\documentclass[10pt,twocolumn]{article}
\usepackage[top=1.9cm, bottom=1.9cm, left=1.8cm, right=1.8cm]{geometry}
\usepackage{times}  %
\usepackage[sort&compress,numbers]{natbib}  %
\usepackage[hyphens]{url}
\usepackage{graphicx}  %
\usepackage[keeplastbox]{flushend}
\usepackage{graphicx} %
\usepackage{tikzsymbols}

\let\oldbibliography\thebibliography
\renewcommand{\thebibliography}[1]{%
\oldbibliography{#1}%
\setlength{\itemsep}{2pt}%
}
\usepackage[hang,flushmargin]{footmisc}

\usepackage[compact]{titlesec}
\titlespacing*{\section}{0pt}{*3}{3pt}
\titlespacing*{\subsection}{0pt}{*2}{2pt}
\usepackage{xspace}
\makeatletter
\def\url@leostyle{%
  \@ifundefined{selectfont}{\def\UrlFont{}}%
  {\def\UrlFont{}}%
}
\makeatother
\usepackage[hyphenbreaks]{breakurl}
\usepackage[bookmarks=true, bookmarksnumbered=true, colorlinks=true, linkcolor=linkcol, citecolor=citecol, urlcolor=urlcol, hypertexnames=true]{hyperref}
\definecolor{darkgreen}{RGB}{0, 100, 0}
\definecolor{linkcol}{rgb}{0.3,0,0}
\definecolor{citecol}{rgb}{0.3,0,0}
\definecolor{urlcol}{rgb}{0.3,0,0}
\definecolor{vlightgray}{gray}{0.925}

\usepackage[small,bf]{caption}
\usepackage{tikz}
\usepackage{amsmath}
\usepackage[sort&compress]{natbib}
\usepackage{comment}
\usepackage{booktabs}
\usepackage{wrapfig}
\usepackage{xspace}
\usepackage{mdframed}
\usepackage{fdsymbol}

\newif\ifcomment
\commentfalse
\ifcomment
\newcommand{\sz}[1]{{\bf \textcolor{brown}{SZ: #1}}}
\newcommand{\jbnote}[1]{{\bf \textcolor{magenta}{JB: #1}}}
\newcommand{\edc}[1]{{\bf \textcolor{blue}{EDC: #1}}}
\newcommand{\gs}[1]{{\bf \textcolor{red}{GS: #1}}}
\newcommand{\hs}[1]{{\bf \textcolor{orange}{HS: #1}}}
\else
\newcommand{\sz}[1]{}
\newcommand{\jbnote}[1]{}
\newcommand{\edc}[1]{}
\newcommand{\gs}[1]{}
\newcommand{\hs}[1]{}
\fi
\newcommand{\descr}[1]{\smallskip\noindent\textbf{#1}}

\usepackage{subfigure}
\makeatletter
\def\url@leostyle{%
  \@ifundefined{selectfont}{\def\UrlFont{\small}}%
  {\def\UrlFont{}}%
}
\makeatother
\newif\ifrevision
\revisionfalse
\ifrevision
\newcommand{\revision}[1]{\textcolor{blue}{#1}}
\newcommand{\revisioncomment}[1]{\textbf{\textcolor{red}{[#1]}}}
\else
\newcommand{\revision}[1]{\textcolor{black}{#1}}
\newcommand{\revisioncomment}[1]{}
\fi

\usepackage{ifthen}
\newif\ifshort
\shorttrue   
\ifshort
	\newcommand{\isShort}{true}
\else
	\newcommand{\isShort}{false}
\fi
\newcommand{\shortVer}[1]{\ifthenelse{\equal{\isShort}{true}}{{#1}}{}}
\newcommand{\longVer}[1]{\ifthenelse{\equal{\isShort}{false}}{{#1}}{}}

\ifshort
\usepackage[keeplastbox]{flushend}
\else
\fi

\begin{document}
\date{}

\title{Understanding and Detecting Hateful Content using Contrastive Learning}

\author{Felipe González-Pizarro$^{\vardiamondsuit}$\thanks{Work done during an internship at Max Planck Institute for Informatics.} , and Savvas Zannettou$^{\spadesuit \blacktriangle}$ \\[0.5ex]
\normalsize $^{\vardiamondsuit}$University of British Columbia, 
\normalsize$^{\spadesuit}$TU Delft, $^{\blacktriangle}$Max Planck Institute for Informatics\\
\normalsize felipegp@cs.ubc.ca, s.zannettou@tudelft.nl}

\maketitle

\begin{abstract}
The spread of hate speech and hateful imagery on the Web is a significant problem that needs to be mitigated to improve our Web experience.
This work contributes to research efforts to detect and understand hateful content on the Web by undertaking a multimodal analysis of Antisemitism and Islamophobia on 4chan's /pol/ using OpenAI's CLIP. This large pre-trained model uses the Contrastive Learning paradigm.
We devise a methodology to identify a set of Antisemitic and Islamophobic hateful textual phrases using Google's Perspective API and manual annotations.
Then, we use OpenAI's CLIP to identify images that are highly similar to our Antisemitic/Islamophobic textual phrases.
By running our methodology on a dataset that includes 66M posts and 5.8M images shared on 4chan's /pol/ for 18 months, we detect 173K posts containing 21K Antisemitic/Islamophobic images and 246K posts that include 420 hateful phrases.
Among other things, we find that we can use OpenAI's CLIP model to detect hateful content with an accuracy score of 0.81 (F1 score = 0.54). \revision{By comparing CLIP with two baselines proposed by the literature, we find that CLIP outperforms them, in terms of accuracy, precision, and F1 score, in detecting Antisemitic/Islamophobic images.} \revisioncomment{MR.3; R1.1} Also, we find that Antisemitic/Islamophobic imagery is shared in 
\revision{a similar number of posts} on 4chan's /pol/ compared to Antisemitic/Islamophobic textual phrases, highlighting the need to design more tools for detecting hateful imagery.
\revision{Finally, we make available (upon request) a dataset of 246K posts containing 420 Antisemitic/Islamophobic phrases and 21K likely Antisemitic/Islamophobic images (automatically detected by CLIP)} \revisioncomment{R3.1} that can assist researchers in further understanding Antisemitism and Islamophobia. 
\end{abstract}


\section{Introduction}
The spread of hateful content on the Web is an everlasting and vital issue that adversely affects society.
The problem of hateful content is longstanding on the Web for various reasons.
First, there is no scientific consensus on what constitutes hateful content (i.e., no definition of what hate speech is)~\cite{sellars2016defining}.
Second, the problem is complex since hateful content can spread across various modalities (e.g.,  text, images, videos, etc.), and we still lack automated techniques to detect hateful content with acceptable and generalizable performance~\cite{arango2019hate}.
Third, we lack moderation tools to proactively prevent the spread of hateful content on the Web~\cite{konikoff2021gatekeepers}.
This work focuses on assisting the community in addressing the issue of the lack of tools to detect hateful content across multiple modalities.

Most of the research efforts in the space of detecting hateful content focus on designing and training machine learning models that are specifically tailored towards detecting specific instances of hateful content (e.g., hate speech on text or particular cases of hateful imagery).
Some examples of such efforts include Google's Perspective API~\cite{jigsaw2018perspective} and the HateSonar classifier~\cite{davidson2017automated} that aim to detect toxic and offensive text.
Other methods aim to detect instances of hateful imagery like Antisemitic images~\cite{zannettou2020quantitative} or hateful memes~\cite{kiela2020hateful,zannettou2018origins}.
These efforts and tools are essential and valuable, however,  they rely on human-annotated datasets that are expensive to create, and therefore they are also small.
At the same time, these datasets focus on specific modalities (i.e., text or images in isolation).
All these drawbacks limit their broad applicability.

The lack of large-scale annotated datasets for solving problems like hate speech motivated the research community to start developing techniques that learn from unlabeled data (a paradigm known as \emph{self-supervised learning}).
Over the past years, the research community released large-scale models that depend on huge unlabeled datasets such as OpenAI's GPT-3~\cite{brown2020language}, Google's BERT~\cite{devlin2018bert}, OpenAI's CLIP~\cite{radford2021learning}, etc.
These models are trained on large-scale datasets and usually can capture general knowledge extracted from the datasets that can be valuable for performing classification tasks that the model was not explicitly trained on.

Motivated by these recent advancements on large-scale pre-trained machine learning models, in this work, we investigate how we can use such models to detect hateful content on the Web across multiple modalities (i.e., text and images).
Specifically, we focus on OpenAI's CLIP model because it helps us capture content similarity across modalities (i.e., measure similarity between text and images).
To achieve this,  CLIP leverages a paradigm known as Contrastive Learning; the main idea is that the model maps text and images to a high-dimensional vector space and is trained in such a way that similar text/images are mapped closer to this vector space (for more details see Section~\ref{sec:background}).



\descr{Focus \& Research Questions.} 
This work focuses on understanding the spread of Antisemitic/Islamophobic content on 4chan's /pol/ board. \revision{We concentrate on hateful content targeted towards these two demographics mainly because previous work indicates that 4chan's /pol/ is known for disseminating Antisemitic/Islamophobic content~\cite{zannettou2020quantitative,prisk2017hyperreality}} \revisioncomment{R2.3}.
Specifically,  we focus on shedding light on the following research questions:

\begin{itemize}
    \item \textbf{RQ1:} Can large pre-trained models that leverage the Contrastive Learning paradigm, like OpenAI's CLIP, identify hateful content with acceptable performance? How does CLIP's performance compare to state-of-the-art classifiers for detecting hateful imagery?
    \item \textbf{RQ2:} How prevalent is Antisemitic/Islamophobic imagery and textual hate speech on 4chan's /pol/? 
   \longVer{\item \textbf{RQ3:} What is the interplay between hateful text and imagery on 4chan's /pol/? For instance, are there specific hateful text instances related to a large volume of hateful images? }
\end{itemize}

To answer these research questions,  we obtain all the posts and images shared on 4chan's /pol/ between July 1, 2016, and December 31, 2017, ultimately collecting 66M textual posts and 5.8M images. 
Then, we leverage the Perspective API and manual annotations to construct a dataset of 420 Antisemitic and Islamophobic textual phrases. We retrieve 246K posts that include any of our 420 hateful phrases. Finally,  we use OpenAI's CLIP to detect Antisemitic/Islamophobic images when provided as input the above-mentioned hateful phrases and all images shared on 4chan's /pol/; we find 21K images that are likely Antisemitic/Islamophobic.

\descr{Contributions \& Main Findings.} Our work makes the following contributions/main findings:
\begin{itemize}
    \item We investigate whether large pre-trained models based on Contrastive Learning can assist in detecting hateful imagery. We find that large pre-trained models like OpenAI's CLIP~\cite{radford2021learning} can detect Antisemitic/Islamophobic imagery with 0.81, 0.54, 0.53, 0.54, accuracy, precision, recall, and F1 score, respectively. \revision{The CLIP model outperforms two baselines that detect hateful imagery in terms of accuracy (0.11 increase), precision (0.17 increase), and F1 score (0.08 increase)} (\textbf{RQ1}) \revisioncomment{MR.3; R1.1}.
    \item We find that on 4chan's /pol/ Antisemitic/Islamophobic imagery appears in a similar number of posts compared to Antisemitic/Islamophobic textual hateful content. This finding highlights the need for the development and use of multimodal hate speech detectors for understanding and mitigating the problem (\textbf{RQ2}).
    \longVer{\item By studying the connection between the Antisemitic/Islamophobic phrases and imagery, we find that Islamophobic imagery is more diverse compared to Antisemitic imagery. That is, we find that 70\% of the Islamophobic phrases are similar to at least 1K images, while for Antisemitism only 54\% of the phrases are similar to at least 1K images (\textbf{RQ3}).}
    \item We make available (upon request from researchers) a large dataset of Antisemitic/Islamophobic posts, phrases, and images shared on 4chan's /pol/.
    \revision{The released dataset includes 892 Antisemitic/Islamophobic images and 420 Antisemitic/Islamophobic phrases that are annotated by the authors of this paper (we expect no false positives in our human-annotated set). Additionally, we release a set of 21K images that were detected by the CLIP model as being Antisemitic/Islamophobic. Note that the set of 21K images includes images that are false positives, since the CLIP model has a precision score of 0.54 for detecting Antisemitic/Islamophobic images.
    Therefore, researchers aiming to use the dataset and create classifiers for Antisemitic/Islamophobic images, should consider the existence of false positives in the dataset.} \revisioncomment{R3.1}
    Nevertheless, we argue that the released dataset can assist researchers in future work focusing on detecting and understanding the spread of hateful content on the Web across multiple modalities (i.e., text and images).
    
\end{itemize}

\descr{Ethical Considerations.} \revision{We emphasize that we rely entirely on publicly available and anonymous data shared on 4chan's /pol/, hence we do not and are unable to obtain consent from users that shared posts/images anonymously on 4chan.
Also, all of our study's manual annotations (i.e., given an image/phrase, annotate if the image/phrase is Antisemitic/Islamophobic) were exclusively performed by the authors of this work.
Given that we only analyze publicly available anonymous data from 4chan and that all manual annotations are undertaken by the authors of this paper, our work is not considered human's subject research from our institution's Ethical Board Committee. 
For our analysis, we follow standard ethical guidelines~\cite{rivers2014ethical} like reporting our results on aggregate and not attempting to deanonymize users.}

\revision{We also discuss some ethical implications with releasing our dataset and possible misuse of the dataset. There is a possibility that malicious actors can make use of our dataset and train  models based on Generative Adversarial Networks~\cite{goodfellow2014generative} with the goal to automatically generate new Antisemitic/Islamophobic imagery. Subsequently, malicious actors can share these Antisemitic/Islamophobic imagery on social media platform, hence affecting people. To minimize this risk, we will make the dataset available only upon request and only to researchers that can provide a description of their intended use.}\revisioncomment{MR.2, MR.5}

\descr{Disclaimer.} This manuscript contains Antisemitic and Islamophobic textual and graphic elements that are offensive and are likely to disturb the reader. 

\section{Background}
\label{sec:background}
This section provides background information on Contrastive Learning and OpenAI's CLIP model, on Google's Perspective API, and 4chan's /pol/ that is our data source.

\descr{Contrastive Learning.} To understand Contrastive Learning, it is essential to grasp its differences compared to traditional Machine/Deep Learning (ML/DL) classifiers.
Traditional ML/DL classifiers take as an input a set of unlabeled data, each accompanied with a class, and aim to predict the class from the unlabeled data, a paradigm known as supervised learning~\cite{caruana2006empirical}.

On the other hand, Contrastive Learning is a self-supervised technique, meaning that there is no need to have classes and models learn from unlabeled data.
The main idea behind Contrastive Learning is that you train a model that relies on unlabeled data, and the model learns general features from the dataset by teaching it which input samples are similar/different to each other~\cite{hadsell2006dimensionality}.
In other words, Contrastive Learning relies on a set of unlabeled data samples with additional information on which of these samples are similar to each other.

Contrastive Learning is becoming increasingly popular in the research community with several applications on visual representations~\cite{chen2020simple,kim2020mixco}, textual representations~\cite{giorgi2020declutr,wu2020clear,gao2021simcse}, graph representations~\cite{you2020graph,hassani2020contrastive}, and multimodal (i.e., text/images, images/videos, etc.) representations~\cite{radford2021learning,diba2021vi2clr,yuan2021multimodal,zhang2020contrastive}.
\begin{figure*}[t]
    \centering
\includegraphics[width=0.8\textwidth]{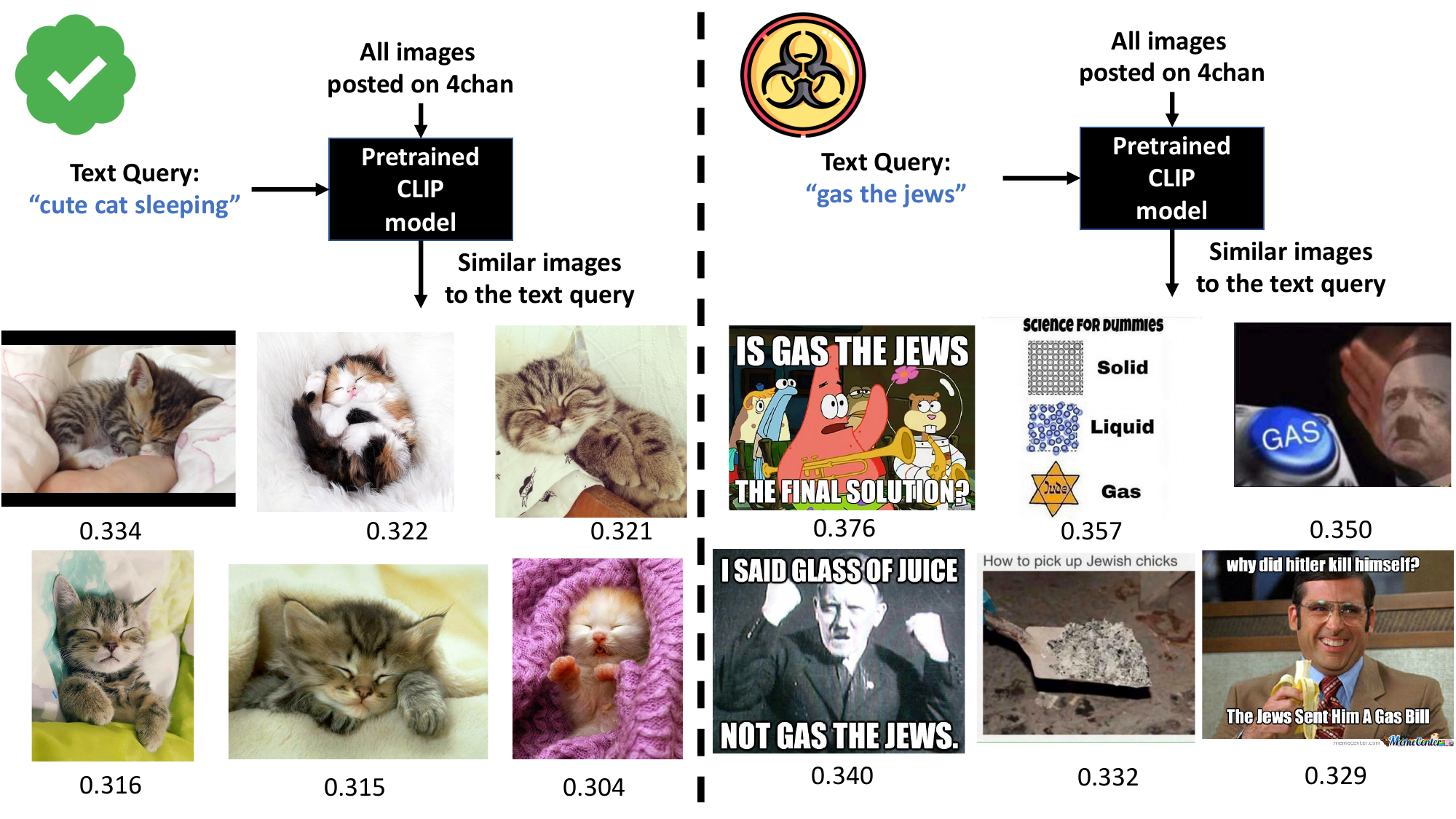} %
\caption{Example of text queries and images that are similar to the text queries on 4chan (i.e., cosine similarity between the text CLIP-representation and the image CLIP-representation equals to 0.3 or more). On the left side, we show a benign text query (``cute cat sleeping''), while on the right we show the results for a toxic and antisemitic query (``gas the jews'') \revisioncomment{R2.7}}
\label{fig:example-clip-4chan}
\end{figure*}

\descr{OpenAI's CLIP.} OpenAI recently released a model called Contrastive Language-Image Pre-training (CLIP)~\cite{radford2021learning} that leverages Contrastive Learning to generate representations across text and images.
The model relies on a text encoder and an image encoder that maps text and images to a high-dimensional vector space. Subsequently, the model is trained to minimize the cosine distance between similar text/image pairs.
To train CLIP, OpenAI created a huge dataset that consists of 400M pairs of text/images collected from various Web sources and covers an extensive set of visual concepts\footnote{The exact methodology for creating this dataset was not made publicly available by OpenAI.}.
By training CLIP with this vast dataset, the model learns general visual representations and how these representations are described using natural language, which results in the model obtaining general knowledge in various topics (e.g., identify persons, objects, etc.).

In this work, we leverage the CLIP model to extract representations for our 4chan textual/image datasets and assess the similarity between the text and image representations.
The main idea is that by providing to the pre-trained CLIP model a set of hateful text phrases, we will be able to identify a set of hateful images that are highly similar to the hateful text query.
\revision{To demonstrate CLIP's potential in discovering hateful imagery from hateful text-based queries, we show an example from our 4chan dataset in Fig.~\ref{fig:example-clip-4chan}.
On the left side, we show examples of images that are highly similar to a benign text query like ``cute cat sleeping,'' while on the right, we show examples of images that are similar to an antisemitic and toxic phrase (``gas the jews'')\footnote{In this work, we treat an image as similar to the text phrase if it has a cosine similarity of 0.3 or higher (see Section~\ref{sec:finding_islamophobic_images}).}.
For each image, we report the cosine similarity between the representation obtained from the text query and the representation of the image in our dataset.
This example shows that the CLIP model can detect objects in images (i.e., cats) and provide relevant images to the queries (i.e., the cats are indeed sleeping according to the query). Furthermore, by looking at the images for the toxic query, we observe that CLIP can identify harmful images based on the query and can link historical persons to it (e.g., the textual input does not mention Adolf Hitler; however, the model knows that Hitler was responsible for the holocaust).
Also, CLIP can detect images that share hateful ideology by adding text on memes (i.e., CLIP also performs Optical Character Recognition and can correlate that text with the text-based query).
Overall, this example shows the predictive power of the CLIP model in detecting hateful imagery from hateful text phrases.} \revisioncomment{R2.6}

\descr{Google's Perspective API.} As a first step towards identifying hateful phrases, we use Google's Perspective API~\cite{jigsaw2018perspective,perspective_library}, which provides a set of Machine Learning models for identifying how rude/aggressive/hateful a comment is.
We use the Perspective API for identifying hateful text mainly because it outperforms other publicly available hate speech classifiers like HateSonar~\cite{davidson2017automated,zannettou2020measuring}.
This work focuses on the SEVERE\_TOXICITY model available from Perspective API because it is more robust to positive uses of curse words \cite{perspective_library}, and it is a production-ready model.
The SEVERE\_TOXICITY model returns a score between 0 and 1, which can be interpreted as the probability of the text being rude and toxic.

\descr{4chan's /pol/.} \revision{4chan is an anonymous image board usually exploited by troll users~\cite{hine2017kek}. On 4chan, users can create a thread by creating a post that contains an image, and other users can create replies with or without images, and they might add references to previous posts. 4chan is well-known for its anonymity and ephemerality. These are the main reasons its users are aggressive in their posts, as there is a lack of accountability~\cite{bernstein2011analysis}. \revisioncomment{MR6, R2.2}}
%
\revision{Our work focuses on 4chan, particularly the Politically Incorrect board (/pol/).
/pol/ is the main board for discussing world events and politics and is known for the spread of conspiracy theories~\cite{zannettou2017web,tuters2018post} and hateful content~\cite{hine2017kek,zannettou2020quantitative}.} \revisioncomment{MR.6, R2.2}

\section{Dataset}

We collect the data about posts on 4chan's /pol/ using the publicly available dataset released by~\citet{papasavva2020raiders}; the dataset includes textual data about 134.5M posts shared on /pol/ between June 2016 and November 2019.
Our work focuses on the period between July 1, 2016, and December 31, 2017 (to match the time period of the image dataset mentioned below), including 66,383,955 posts.
We complement the above dataset with the image dataset collected by~\citet{zannettou2020quantitative}.
The dataset includes 5,859,439 images shared alongside /pol/ posts between July 1, 2016, and December 31, 2017.
Overall, our dataset comprises all textual and image activity on /pol/ between July 1, 2016, and December 31, 2017, including 66M posts and 5.8M images.


\section{Methodology}


This section describes our methodology for detecting hateful text phrases and hateful imagery, focusing on Antisemitic and Islamophobic content. 



\subsection{Identifying Antisemitic and Islamophobic phrases \revisioncomment{R2.4}}

Here, our goal is to identify a set of phrases that are Antisemitic/Islamophobic.
To do this, we follow a multi-step semi-automated methodology.
First, we use the SEVERE\_TOXICITY scores from the Perspective API to identify posts that are toxic/offensive without considering the target (e.g., if it is antisemitic). 
Specifically, we consider all posts that have a score of 0.8 or more as toxic, following the methodology by~\citet{ribeiro2020does}.
Out of the 66M posts in our dataset, we find 4.5M (6.7\%) toxic posts.

Having extracted a set of toxic posts from 4chan's /pol/, we then aim to identify the main targets of hate speech on /pol/ by extracting the top keywords.
To do this, we preprocess the data to remove HTML tags, stop words, and URLs, and then we create a term frequency-inverse document frequency array (TF-IDF). Next, we manually inspect the top 200 words based on their TF-IDF values and identify the words related to Jews or Muslims.
As a result, we find seven keywords: ``jews,'' ``kike,'' ``jew,'' ``kikes,'' ``jewish,'' ``muslims,'' and ``muslim.''
Then, based on these keywords, we filter the toxic posts obtained from the previous step, hence getting a set of 336K posts with a SEVERE\_TOXICITY score of 0.8 and include at least one of the seven keywords. \revision{Note that we decide to focus on the top 200 terms because we want to focus on popular targeted groups in the set of toxic posts.} \revisioncomment{R2.1, R2.10} 
\revision{Moreover, while terms such as ``Islam'' or ``Islamic'' might appear crucial to include in our list of keywords, we not include them because their TF-IDF scores are far from the top ones. Indeed, the  term ``islam'' occupies the position 275th, ``islamic'' the 1014th position, ``mohammed'' the position 1553th, and ``prophet'' the position 2505th.} \revisioncomment{R2.11}.

\revision{Since our goal is to create a set of Antisemitic/Islamophobic phrases, we need to break down the toxic 4chan posts into sentences and then identify the ones that are Antisemitic/Islamophobic.
To do this, we apply a sentence tokenizer~\cite{sentence_tokenizer} on the 336K posts, obtaining 976K sentences.
To identify common phrases used on 4chan's /pol/, we apply WordNet lemmatization~\cite{wordnet}, excluding all sentences that appear less than five times.} \revisioncomment{R2.12} We obtain 4,582 unique common phrases; not all of these sentences are Antisemitic/Islamophobic. \revision{We note that some phrases are contained in longer phrases. However, we do not treat them as duplicates, given that the text encoder of the CLIP model encodes them differently.}\revisioncomment{R2.16}

\begin{table}[t]
\centering
\small
\begin{tabular}{@{}lrr|rr@{}}
\toprule
\textbf{Dataset}      & \multicolumn{2}{c|}{\textbf{Textual}}                                                                                   & \multicolumn{2}{c}{\textbf{Visual}}                                                                                   \\ \midrule
                      & \multicolumn{1}{l}{\textbf{\# Phrases}} &  \multicolumn{1}{l|}{\textbf{\# Posts}} & \multicolumn{1}{l}{\textbf{\# Images}} &  \multicolumn{1}{l}{\textbf{\# Posts}} \\ \midrule
\textbf{Antisemitism} & 326                                                             & 209,224                           & 15,711                                                            & 143,506                              \\

\textbf{Islamophobia} & 94                                                                   & 37,354                                & 5,548                                                                  & 29,978                              \\ \midrule
\textbf{Total}        &  420                                                                      & 246,578                                    & 21,259                                                                   & 173,484                                \\ \bottomrule
\end{tabular}
\caption{\revision{Overview of our Antisemitism/Islamophobia Textual and Visual datasets. The number of phrases is based on lemmatized versions, and the number of images is based on the unique pHash values. We consider only images associated with at least 10 phrases to reduce the number of false positives.} \revisioncomment{R2.6} }
\label{tab:number_posts_both_datasets_images_filtered}
\end{table}

Identifying whether a phrase is Antisemitic/Islamophobic is not a straightforward task and can not be easily automated. Therefore, we use manual annotation on the 4,582 common phrases to annotate the common phrases as Antisemitic/Islamophobic or irrelevant.
Two authors of this paper independently annotated the 4.5K common phrases. \revision{On average, these phrases include 11.10 words ($\sigma$ = 20.82).}\revisioncomment{R2.13} We discard long phrases (over seven words) during the annotation since our preliminary experiments showed that OpenAI's CLIP returns a considerable amount of false positives when provided with long text queries. We also consider as irrelevant phrases that target multiple demographic groups (e.g., hateful towards Muslims and Jews like ``fuck jews and muslims'' or hateful towards African Americans and Jews like ``fuck niggers and jews''). \revision{To ease the annotation process, we create a spreadsheet that includes a clear description of our labels, as well as all the 
information that an annotator needs in order to inspect and correctly annotate a phrase (e.g., number of terms per phrase). Phrases labelled as Antisemitic express hostility to, prejudice towards, or discrimination against Jews~\cite{antisemitism_definition}. Phrases labelled as Islamophobic express fear of, hatred of, or prejudice against the Islam or Muslims in general~\cite{islamophobia_definition}.
The two annotators agreed on 91\% of the annotations with a Cohen's Kappa score of 0.69, which indicates a substantial agreement~\cite{kappa}.
After the independent annotations, the two annotators discussed the disagreements to come up with a final annotation on whether a phrase is Antisemitic/Islamophobic or irrelevant.
After our annotation, we find 326 Antisemitic and 94 Islamophobic phrases.
The list of the Antisemitic/Islamophobic phrases is publicly available~\cite{url_toxic_phrases}} \revisioncomment{MR.6}. 

\revision{Finally, we search for these Antisemitic/Islamophobic phrases on the entire dataset. We extract all posts that include any of the Antisemitic/Islamophobic phrases (\emph{Textual} dataset), finding 247K posts. Note that we remove 864 (0.35\%) posts that contain both Antisemitic and Islamophobic phrases. Overall, we find 209K (84.85\%) Antisemitic posts and 37K (15.15\%) Islamophobic posts (see Table~\ref{tab:number_posts_both_datasets_images_filtered})} \revisioncomment{R2.14}. 

\subsection{Identifying Antisemitic and Islamophobic images}
\label{sec:finding_islamophobic_images}


Our goal is to identify Antisemitic and Islamophobic imagery using the pre-trained CLIP model~\cite{radford2021learning}.
To do this, we encode all images in our dataset using the image encoder on the CLIP model, hence obtaining a high-dimensional vector for each image.
Also, we encode all the Antisemitic/Islamophobic phrases (extracted from the previous step), using the text encoder on the CLIP model, obtaining a vector for each phrase.
Then, we calculate all the cosine similarities between the image and text vectors, which allows us to assess the similarity between the phrases and the images.
The main idea is that by comparing a hateful phrase to all the images, images with a high cosine similarity score will also be hateful.
To identify a suitable cosine similarity threshold where we treat a text and an image similarly, we perform a manual annotation process.

\begin{figure}[t!]
\centering
\includegraphics[width=0.8\columnwidth]{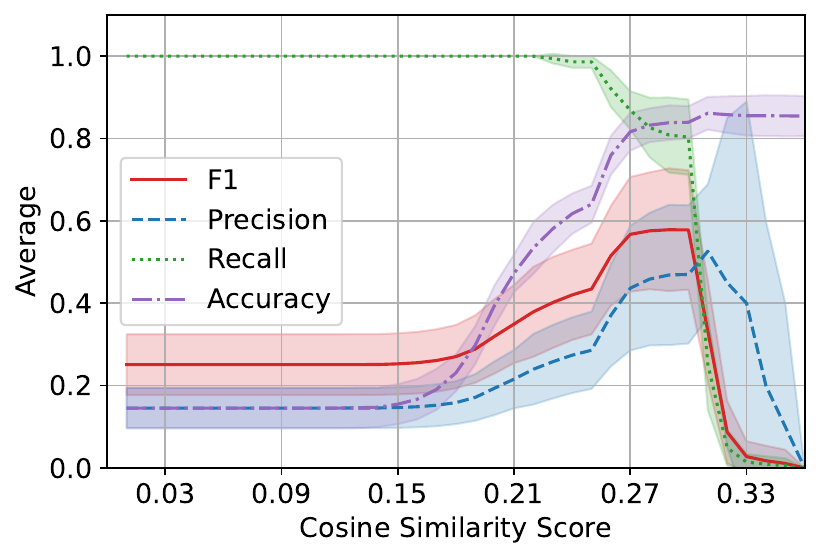}
\caption{Performance of the CLIP model in identifying Antisemitic/Islamophobic imagery for varying cosine similarity thresholds. The lines refer to the average metric for ten random phrases (2K images), while the area refers to the standard deviation across the ten phrases.}
\label{fig:metrics_image_annotation}
\end{figure}

\descr{Identifying a suitable threshold.} \revision{First, we extract a random sample of ten Antisemitic/Islamophobic phrases (eight Antisemitic and two Islamophobic to match the percentage of Antisemitic/Islamophobic phrases in our dataset). 
Then, we extract a random sample of 200 images for each phrase while ensuring that the images cover the whole spectrum of cosine similarity scores.
Specifically, we extract 50 random images with cosine similarity scores for each of the following ranges: [0.0, 0.20), [0.2, 0.25), [0.25, 0.3), [0.3, 0.4]. 
To select these ranges, we plot the} \revision{Cumulative Distribution Function (CDF) }\revision{ of all cosine similarity scores obtained by comparing the ten randomly selected phrases and all the images in our dataset (we omit the figure due to space constraints).
We find that 40\% of the scores are below 0.2, and we expect these images to be entirely irrelevant for the phrase.
To verify this, we select the [0.0, 0.20) range.
Additionally, we select the [0.2, 0.25] because it has a considerable percentage of the scores (50\%), and we expect that the images will not be very similar again.
Finally, we select the [0.25, 0.3) and [0.3-0.4] ranges because we expect that the ideal threshold is somewhere in these two ranges, and devoting half of the selected images in these ranges will help us identify a suitable threshold.} \revisioncomment{R2.8, R2.17}

\revision{Then, two authors of this paper independently annotated the 2,000 images to identify which are Antisemitic/Islamophobic or irrelevant.
In a similar fashion to our annotations for toxic phrases, we labeled images as Antisemitic, those that clearly express hostility, prejudice towards, or discrimination against Jews~\cite{antisemitism_definition}. Images labeled as Islamophobic clearly express hatred of or prejudice against Muslims~\cite{islamophobia_definition}.
The annotators agreed on 94\% of the annotations with a Cohen's Kappa score of 0.75, which indicates a substantial agreement. Again, the two annotators solved the disagreements by discussing the images and deciding a final annotation on whether the image is Antisemitic/Islamophobic or irrelevant.} \revisioncomment{MR.6, R2.8}


\revision{As a result, our initial ground truth dataset of Antisemitic and Islamophobic imagery includes 291 (14.55\%) hateful images: 239 (82.13\%) of them are Antisemitic and 52 (17.87\%) are Islamophobic.}\revisioncomment{R2.18} Having constructed an initial ground truth dataset of Antisemitic and Islamophobic imagery, we then find the best performing cosine similarity threshold.
We vary the cosine similarity threshold, and we treat each image as Antisemitic/Islamophobic (depending on the phrase used for the comparison) if the cosine similarity between the phrase and the image is above the threshold.
\revision{Then, we calculate the accuracy, precision, recall, and F1 score, for each of the ten phrases.
We report the average performance across all phrases and the standard deviation (as the area) in Fig.~\ref{fig:metrics_image_annotation}.
We observe that the model performs best with a cosine similarity threshold of 0.3 as we achieve a 0.84, 0.47, 0.80, 0.58 for accuracy, precision, recall, and F1 score, respectively.} \revisioncomment{R2.8}
Indeed, the 0.3 threshold is also used by previous work by~\citet{schuhmann2021laion} that inspected CLIP's cosine similarities between text and images and determined that 0.3 is a suitable threshold.


To construct our initial Antisemitic/Islamophobic image dataset, we extract all images that have a cosine similarity of 0.3 or higher with any of the Antisemitic/Islamophobic text phrases.
We label each image as likely Antisemitic or likely Islamophobic depending on whether the textual phrase is Antisemitic or Islamophobic. 
To identify unique images, we use the Perceptual Hashing (pHash) algorithm~\cite{Monga2006PerceptualIH} that calculates a fingerprint for each image in such a way that any two images that look similar to the human eye map have minor differences in their hashes. Similar to the Textual dataset, we remove all images labeled as both Antisemitic and Islamophobic (3,325 images), mainly because our manual inspections indicate that most of them are noise.
Overall, we find 69,610 likely Antisemitic and 22,519 likely Islamophobic images that are shared in 472,048 and 101,465 posts, respectively.


\descr{Evaluating performance to entire dataset.} \revision{To evaluate the quality of our Antisemitic/Islamophobic detection approach in the entire dataset (and not limited to a few phrases as before), we perform an additional manual annotation on 2,000 randomly selected images (from our 92K likely Antisemitic/Islamophobic images mentioned above). 
We obtain 1,507 (75.4\%) potential Antisemitic images and 493 (24.7\%) potential Islamophobic images. Two authors of this paper independently annotated these images to identify which are actually Antisemitic/Islamophobic.} \revisioncomment{R2.6, MR.6} \revision{The two annotators agreed on 84.9\% of the annotations with a Cohen’s Kappa score of 0.64, which indicates a substantial agreement~\cite{kappa}. The annotators identify 551 (27.6\% out of total annotated images) Antisemitic/Islamophobic images: 389 (70.6\%)  are identified as Antisemitic and 162 (29.4\%) are identified as Islamophobic.}\revisioncomment{R2.6}


\begin{figure}[t!]
\centering
\includegraphics[width=0.8\columnwidth]{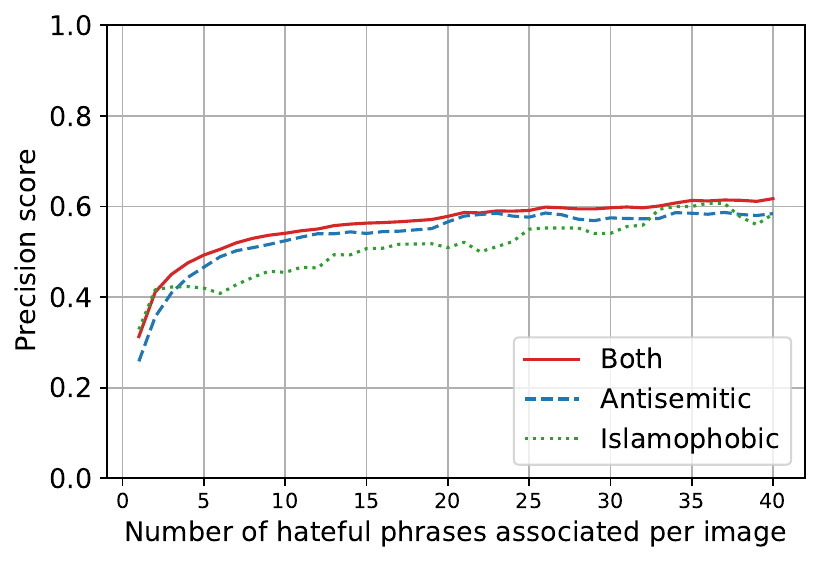}
\caption{\revision{Precision of the CLIP model in identifying Antisemitic/Islamophobic imagery by varying the number of hateful phrases that have a cosine similarity of 0.3 or more with each image.} \revisioncomment{R2.6}}
\label{fig:precision_different_number_queries_per_image}
\end{figure}

\descr{Improving performance.} \revision{Given the relatively small percentage (27.6\%) of the images that are actually Antisemitic/Islamophobic, we set out to investigate how we can improve this performance.
We hypothesize that we can improve the detection performance by considering the number of hateful phrases that have high cosine similarity with the detected images.
Indeed, based on our annotated dataset, we find that images associated with a higher number of hateful phrases are more likely to be Antisemitic/Islamophobic (see Fig. \ref{fig:precision_different_number_queries_per_image}). 
For instance, by considering only images associated (i.e., cosine similarity between the phrase and the image at least 0.3) with ten or more hateful phrases, 54,1\% of them can be identified as Antisemitic/Islamophobic (see Fig. \ref{fig:precision_different_number_queries_per_image}). 
Considering this threshold, the CLIP model has a precision score of 0.57 when identifying Antisemitic imagery and a precision score of 0.43 when identifying Islamophobic imagery. 
For the rest of the analysis, we use this threshold as it greatly reduces the number of false positives that are generated.
Table \ref{tab:number_posts_both_datasets_images_filtered} shows the final number of posts in our Visual dataset. Our final visual dataset contains 21K likely Antisemitic/Islamophobic images.} \revisioncomment{R2.6}


\begin{table}[t]
\centering
\resizebox{\columnwidth}{!}{
\begin{tabular}{@{}lrrrr@{}}
\toprule
                                                                                                                                                
                      & \multicolumn{1}{l}{\textbf{ Accuracy}} &  \multicolumn{1}{l}{\textbf{ Precision}} & \multicolumn{1}{l}{\textbf{ Recall}} &  \multicolumn{1}{l}{\textbf{ F1}} \\ \midrule
MMBT-Grid  & 0.70 & 0.37  &0.61 &0.46  \\
MOMENTA-C  w/o ORC & 0.56 & 0.27 & 0.63 & 0.38  \\
MOMENTA-C & 0.60 & 0.27 & 0.51  & 0.35 \\
MOMENTA-P w/o  ORC & 0.46 & 0.24 & \textbf{0.73} & 0.36  \\
MOMENTA-P &  0.57& 0.29 & 0.69  & 0.40  \\
\midrule
\textbf{CLIP Model}        & \textbf{0.81} & \textbf{0.54} & 0.53 & \textbf{0.54} \\ \bottomrule
\end{tabular}
}
\caption{Performance comparison between CLIP model and the two baselines.  MOMENTA-C and MOMENTA-P corresponds to the MOMENTA model pre-trained on a COVID-19 and US Politics dataset respectively.} 

\label{tab:baseline_models}
\end{table}

\descr{Distance Metric \& Dimensionality.} \revision{We also investigate other ways to improve performance by using different distance metrics or applying dimensionality reduction techniques.
In particular, we experiment with Euclidean distance and Mahalanobis distance~\cite{mahalanobis1936generalized}, with both performing substantially worse compared to cosine distance.
Also, we try reducing the dimensionality of the CLIP embeddings to 64, 128, 256 dimensions using the Uniform Manifold Approximation and Projection approach~\cite{mcinnes2018umap}, without any performance gains.
Note, that we do not include the actual performance with different distance metrics and after dimensionality reduction due to space constraints.
Based on these results, for our detection and analysis, we use the cosine distance metric on the original embeddings obtained from the CLIP model.
}\revisioncomment{MR.7, R3.2}

\descr{Baseline models.} 
\revision{Here, we aim to compare the performance of CLIP model (that considers an image as Antisemitic/Islamophobic if it has a cosine similarity of 0.3 or more for at least 10 hateful phrases), using our final ground truth dataset, which combines the two above-mentioned annotation procedures (4K images).
Our ground truth dataset includes 678 Antisemitic images, 214 Islamophobic images, and 3,158 non-hateful images. 
We compare our method of identifying Antisemitic/Islamophobic imagery with two hateful detection models (see Table \ref{tab:baseline_models}). 
First, \textbf{MMBT-Grid}~\cite{kiela2019supervised} is a multimodal architecture which consists of supervised multimodal transformers using Image-Grid features. We use the pre-trained weights released by~\citet{kiela2020hateful} for hateful image detection. Second, we use \textbf{MOMENTA}~\cite{pramanick-etal-2021-momenta-multimodal}, which analyzes the input's local and global perspective for detecting harmful memes and their targets. Several experiments show that it outperforms several robust approaches.} 
\revision{\textbf{MOMENTA} can identify images that have the potential to cause harm to individuals, organizations, and communities, which is also the focus of our work. This model is pre-trained with two datasets related to COVID-19 and US politics. While our 4chan dataset is different from those, we use MOMENTA as it is a generalizable model~\cite{pramanick-etal-2021-momenta-multimodal}. }\revisioncomment{MR.3; R1.1}

\revision{Table \ref{tab:baseline_models} shows the results for Antisemitic/Islamophobic imagery detection. 
We observe that the CLIP model outperforms all baselines in terms of accuracy, precision, and F1 score, with an improvement of 0.11, 0.17, and 0.08, respectively (compared to the second-best performing model). Also, we find that the CLIP model has a lower recall score compared to the baselines.
Nevertheless, in this work, we favor precision over recall, as we aim to reduce the number of false positives that are generated by our method.} \revisioncomment{MR.3; R.1.1}

\section{Results}
This section presents our results from analyzing the Antisemitic/Islamophobic Textual and Visual datasets.


\begin{figure}[t]
\centering
\includegraphics[width=0.8\columnwidth]{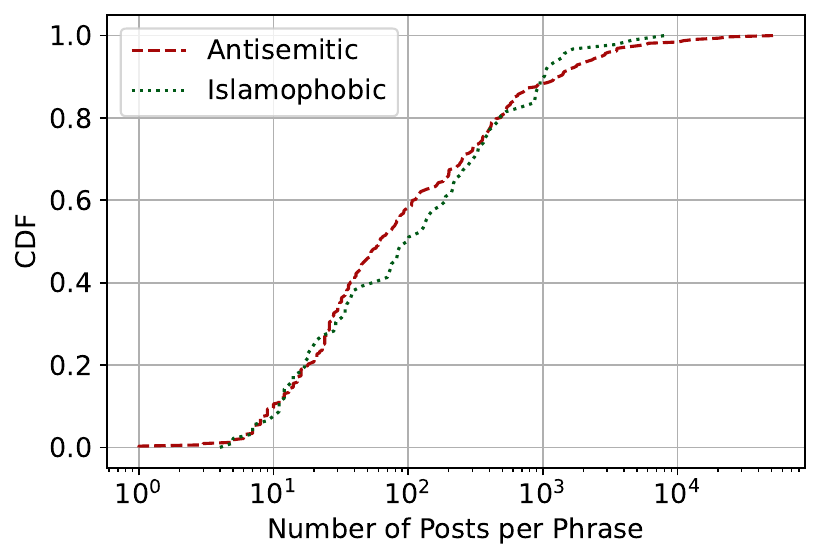}
\caption{CDF of the number of Antisemitic/Islamophobic posts containing each phrase.}
\label{fig:cdf_number_posts_per_phrase}
\end{figure}

\begin{table}[t]
\centering
\resizebox{\columnwidth}{!}{%
\setlength{\tabcolsep}{3pt}
\begin{tabular}{lr|lr}
\toprule
  \multicolumn{2}{c|}{\textbf{Antisemitic phrases}}                                                                                   & \multicolumn{2}{c}{\textbf{Islamophobic phrases}}                                                                                   \\ \midrule
          \multicolumn{1}{l}{\textbf{Phrase}}            & \multicolumn{1}{r|}{\textbf{\# Posts}} & \multicolumn{1}{l}{\textbf{Phrase}} & \multicolumn{1}{r}{\textbf{\# Posts}}  \\ 
          
          \midrule

a kike &	51,216& fuck muslim	&7,993\\
fuck kike	&23,604& kill muslim	&4,639\\
fuck jew	&20,241& fuck islam	&3,464\\
gas the kike	&13,353& kill all muslim	&1,672\\
fuck off kike	&11,108& muslim be terrorist	&1,445\\
kike shill	&10,134& i hate muslim	&1,379\\
gas the kike race war now	&6,105& muslim shithole	&1,208\\
kill jew	&5,815 &muslim shit	&1,085\\
you fuck kike	&5,007& all muslim be terrorist	&1,039\\
filthy kike	&4,111 &muslim be bad	&1032\\
jew fuck	&3,557 &ban all muslim	&958\\
kike faggot	&3,540 &fuck mudslimes	&951\\
kike on a stick	&3,537& muslim cunt	&907\\
gas the jew	&3,081 &i hate islam	&881\\
faggot kike	&2,905 &fuck sandniggers	&876 \\
\bottomrule
\end{tabular}
}
\caption{Top 15 phrases (lemmatized versions), in terms of the number of posts, in our Antisemitic and Islamophobic Textual dataset. For each phrase, we report the number of posts that contain it.}
\label{tab:phrases_and_number_posts_textual_dataset}
\end{table}


\subsection{Popular phrases in Textual dataset}

We start our analysis by looking into the most popular phrases in our Antisemitic/Islamophobic textual datasets. Fig.~\ref{fig:cdf_number_posts_per_phrase} shows the Cumulative Distribution Function (CDF) of the number of posts per each Antisemitic/Islamophobic phrase. We observe that these hateful phrases tend to appear in a considerable amount of posts. For instance, 90.4\%  and 92.47\% of the Antisemitic and Islamophobic phrases appear in at least ten posts. Furthermore, we identify that the percentage of Antisemitic phrases (41.8\%) that appear in at least 100 posts is slightly lower than the percentage of Islamophobic phrases (49.46\%). At the same time, we observe that a small percentage of phrases (11.54\%) is shared in more than 1000 posts on the Antisemitic/Islamophobic textual datasets combined.


We also report the top 15 phrases, in terms of the number of posts, in our Antisemitic and Islamophobic Textual dataset (see Table~\ref{tab:phrases_and_number_posts_textual_dataset}). In the first dataset, we observe that 12 out of the 15 most frequent phrases contain the term ``kike'', a derogatory term to denote Jews. We also identify three phrases related to the extermination procedure in the gas chambers during the holocaust. Indeed, 16,433 (7.85\%) of the Antisemitic posts contain at least one of these phrases: ``gas the kike'', ``gas the jew'', or ``gas the kike race war now''. Phrases accusing jews of being accomplices (``kike shill'') or alluding to a supposed good social-economic status (``filthy kike'') are also trendy, appearing in 10,134 and 4,111 posts, respectively.

We also show the top 15 most popular Islamophobic phrases in Table~\ref{tab:phrases_and_number_posts_textual_dataset}. Here,  we observe many posts with phrases calling Muslims as terrorists. For instance, ``Muslims be terrorist'' and ``All Muslim be terrorist'' appear in 1,445 and 1,039 posts, respectively. The second and fourth most popular phrases are calls for attacks targeting Muslims; ``Kill Muslim'' and ``Kill all Muslim'' appear in approximately 4.6K and 1.6K posts. We also find phrases against Islam; ``Fuck Islam'' appears in 3.4K posts and ``I hate Islam'' in 881 posts. Finally, we also identify phrases containing the terms ``mudslimes''~\cite{mudslime} and ``sandniggers''~\cite{sandnigger}, which are derogatory names to refer to Muslims and Arabs.

\begin{figure}[t!]
\centering
\includegraphics[width=0.8\columnwidth]{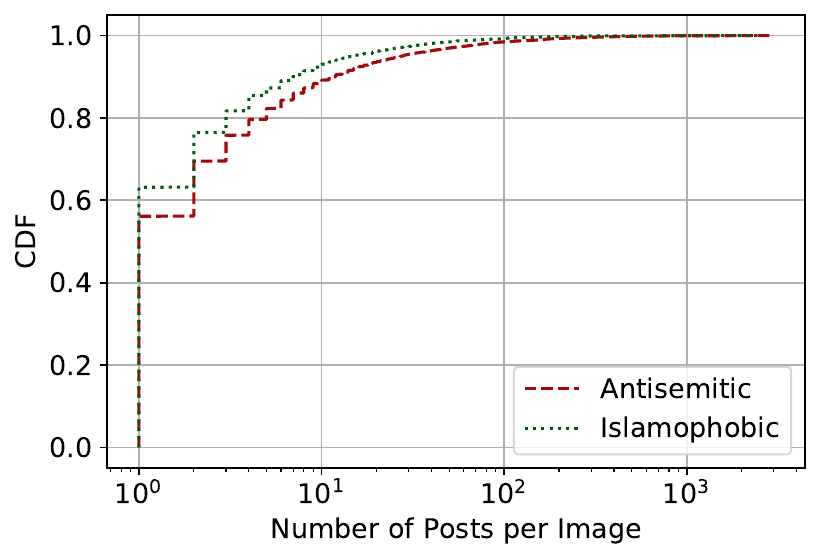}
\caption{CDF of the number of Antisemitic/Islamophobic posts containing each image. }
\label{fig:cdf_number_posts_per_image}
\end{figure}





\begin{figure*}[!h]
    \centering
\includegraphics[width=\textwidth]{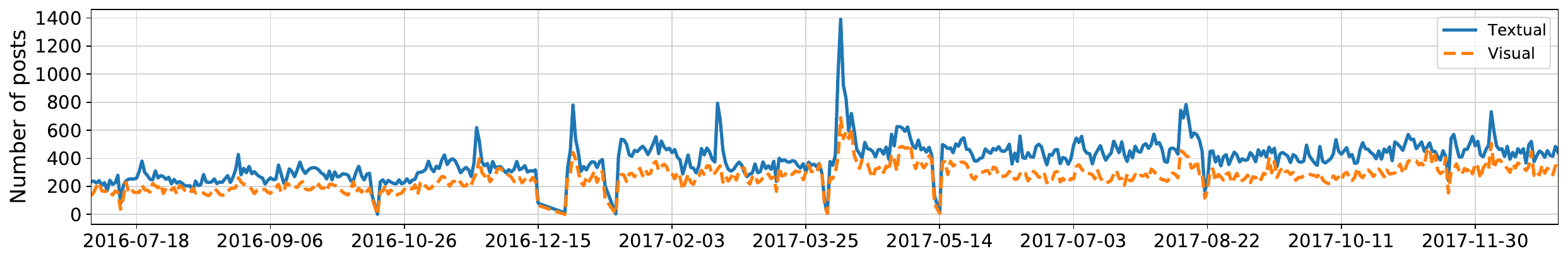} 
\caption{Number of Antisemitic posts per day in our Textual/Visual datasets.} 
\label{fig:antisemitic_overtime}
\end{figure*}

\begin{figure*}[!h]
    \centering
\includegraphics[width=\textwidth]{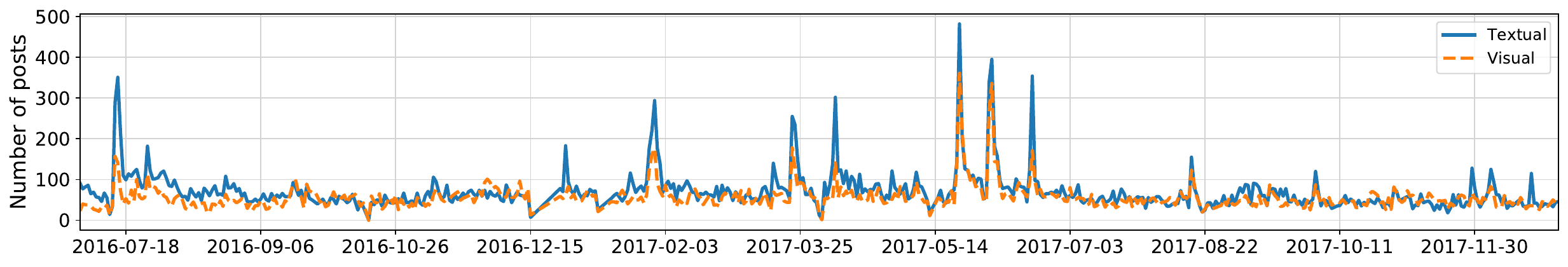} 
\caption{Number of Islamophobic posts per day in our Textual/Visual datasets.}
\label{fig:islamophobic_overtime}
\end{figure*}

\subsection{Popular Images in Visual Dataset}

We also look into the popularity of images in our Antisemitic/Islamophobic datasets (in terms of the number of posts they shared).
Fig.~\ref{fig:cdf_number_posts_per_image} shows the CDF of the number of posts for each Antisemitic/Islamophobic image.
We observe that Antisemitism and Islamophobia imagery is a diverse problem, with 56.13\% and 63.16\% of the images appearing only in one post for Antisemitism and Islamophobia, respectively.
At the same time, we have a small percentage of images that are shared many times on 4chan's /pol/; 10.62\% of all the Antisemitic/Islamophobic imagery are shared at least ten times. Overall, we observe a similar pattern in the distribution of the number of posts per image for Antisemitism and Islamophobia.


\revision{Next, we look into the most popular images  in our Antisemitic and Islamophobic visual datasets. We avoid showing the images since they are highly offensive and are likely to disturb the readers, and only discuss the main insights from inspecting the most popular images. We identify the top ten Antisemitic images. We find that the Happy Merchant meme appears in five out of the top ten Antisemitic images. 
Other interesting examples of popular images are hinting that members of the Jewish are allegedly the masterminds of lousy stuff happening or conspiracy theories (i.e., shut down the Jewish plan or Rabbi painting Nazi symbols). 
We also find two false positives among the top ten Antisemitic images; the one shows Pepe the Frog wearing a t-shirt with a swastika symbol, while the other shows again Pepe the frog dressed as a crusader with the text ``KEK WILLS IT." }

\revision{
We also look at the top ten most popular Islamophobic images. 
We find two images that are pretty graphic and insult Prophet Muhammad and the Holy Quran (again we do not include the images since they are highly offensive).
These two Islamophobic images are included in 1.1K posts within 4chan's /pol/, indicating that graphic images that insult Islam as a religion are used a lot on 4chan. 
Moreover, we find 
images that include sarcasm and link Muslims to terrorism; for instance, CLIP links the phrase ``muslims shihole'' with an image of a Muslim dressed as a terrorist, likely indicating that the CLIP model thinks that Muslims are terrorists.
We also find an image linking Muslims to the Happy Merchant meme; i.e., the Happy Merchant dressed as a Muslim.
Among the top 10 Islamophobic images, we find one image that is a false positive. This image is showing a meme that compares Americans to Europeans and is likely considered as related  because it includes the word ``Muhammad,'' however upon manual examination we do not find this image Islamophobic.}
\revisioncomment{R2.19, M3.1, MR.4}

\subsection{Antisemitic/Islamophobic content over time}

This section presents our temporal analysis that shows the distribution of Antisemitic/Islamophobic content over time.
Fig.~\ref{fig:antisemitic_overtime} shows the number of hateful posts per day in the Antisemitic Textual/Visual datasets.
We run Kendall's tau-b correlation to determine the relationship between the number of posts in the Antisemitic Textual and Visual dataset.
We find a  strong, positive, and statistically significant correlation ($\tau=.635,p<.001)$, indicating that Antisemitic content is spread both using text and images in a similar fashion.
We also observe the highest volume of textual and image content between April 6, 2017, and April 9, 2017, with 4,132 (1.97\% of the dataset) posts in the Textual dataset and 2,223  (1.55\%) posts in the Visual dataset. 
This finding confirms previous findings from~\citet{zannettou2020quantitative} that identified a spike in the spread of the Happy Merchant memes on April 7, 2017. 

By inspecting the top 15 most frequent images during that period (we omit the figure due to space constraints), we identified that those images are related to the decision of Donald Trump to remove Steve Bannon from the National Security Council Post on April 5, 2017 ~\cite{costa2017stephen} and a missile attack in Syria on April 7, 2017 ~\cite{rosenfeld2017trump}. 
According to newspapers ~\cite{baker2017trump,haberman2017battle}, Jared Kushner, the Jewish Trump's son-in-law, seemed to be acting as a shadow secretary of state visiting and taking Middle East portfolios after that event. This political decision spread a volume of image content with the face of Jared Kushner. Also, there are some references to Donald Trump that indicate that he is controlled by Israel (e.g., most popular images are associated with the phrases ``fuck trump and fuck jews'',  and ``fuck trumpstein and fuck jewish people''). 



We also evaluate the distribution of Islamophobic posts over time. Fig.~\ref{fig:islamophobic_overtime} shows the number of Islamophobic posts per day in our Textual/Visual datasets. We also find a statistically significant,  strong, and positive correlation ($\tau=.393,p<.001)$. In both datasets, we find a peak of activity on May 23, 2017, with 482 and 361 posts in the Textual and Visual datasets, respectively.
By manually inspecting the top 15 images shared that day, we identify that the high volume of posts is related to the Manchester Bombing; on May 22, 2017, a British man detonated a suicide bomb in the foyer of the Manchester Arena as people were leaving a concert by pop singer Ariana Grande. On May 23, ISIS claimed responsibility for the attack. \cite{cobain2017salman}. This event raised hateful online narratives defining Muslims as terrorists \cite{downing2022tweeting}. We find images that contain explicit references to this attack and images questioning whether Islam is a religion of peace. 
Overall, our findings highlight that both textual and visual hateful content is likely influenced by real-world events, with peaks of hateful activity observed during important real-world events that are related to the demographic groups we study.

\longVer{
\subsection{Interplay Between Antisemitic/Islamophobic Phrases and Imagery}
We aim to identify how prevalent Antisemitic/Islamophobic imagery is on 4chan's /pol/ and how it compares to the spread of hateful textual phrases. To do so, we calculate the ratio of the number of posts in Textual and Visual datasets for Antisemitic and Islamophobic phrases (number of posts in Textual dataset divided by the number of posts that include similar images for each phrase). 
Fig.~\ref{fig:ratio_per_phrase} shows a CDF of the calculated ratio for each phrase.
We observe that for a  small percentage (7.12\%) of Antisemitic phrases, the textual content is at least ten times higher than the image content. In the case of Islamophobic phrases, there is no phrase with a ratio higher or equal than ten. On the other hand,  our results show that for 43.65\% of the Antisemitic phrases and 40.86\% of the Islamophobic phrases, the number of posts in the Visual dataset is at least 100 times the number of posts in the Textual dataset. Finally, for 17.65\% of the Antisemitic phrases and 6.45\% of the Islamophobic phrases, the number of posts in the Visual dataset is 1000 times the number of posts in the Textual dataset.
This highlights that for a substantial percentage of phrases a large number of images can be used to describe similar Antisemitic/Islamophobic sentiments, which even overshadows the spread of the textual claim on 4chan's /pol/.



\begin{figure}[t!]
\centering
\includegraphics[width=0.8\columnwidth]{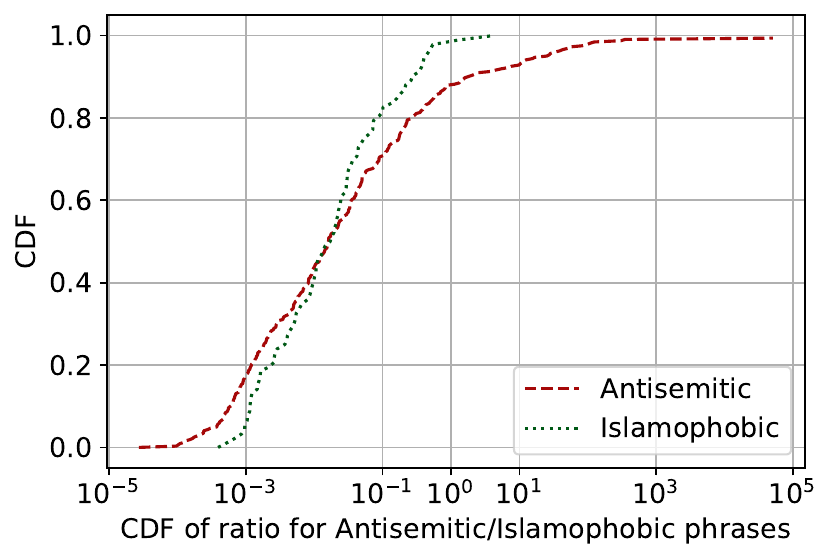} 
\caption{CDF of the ratio in Antisemitic and Islamophobic posts}
\label{fig:ratio_per_phrase}
\end{figure}

Table~\ref{tab:relevant_phrases_ratio_textual_visual_posts} supplements our results by showing the top phrases with the highest and lowest ratio between the number of Textual and Visual posts. We observe that the top five Antisemitic phrases with the highest ratio contain the term ``kike,'' which, as we describe above, is a derogatory term used to offend Jewish people. The number of images related to these phrases is none or just a few. On the other hand, we observe that the number of posts related to the phrases ``Hitler didn't kill enough Jews'' or ``Jewish child should be kill''  is at least 10,000x lower in the Textual dataset than the Visual dataset. These results suggest a considerable difference in the amount of data retrieved for the exact phrase between the Textual and Visual datasets. 
Some of the Islamophobic phrases with the highest ratio also contain derogatory terms against communities. As we described before, the terms ``mudslimes'' and ``sandniggers'' are used to offend Muslims and Arabs, respectively. On the other hand, we find that the number of posts in the Visual dataset is at least 1000 times the number of posts in the Textual dataset for phrases such as ``fuck Muhammad and fuck Muslim'', and ``fuck Islam and fuck all Muslim''.
Overall, we find that the pre-trained CLIP model can detect hateful images more easily from text that does not include slurs like ``kike,'' ``mudslime,'' and ``sandniggers'' likely because the training data did not have many instances of such slur words.

\begin{table*}[t]
\centering
\resizebox{\textwidth}{!}{%
\begin{tabular}{@{}lrrr|lrrr@{}}
\toprule
  \multicolumn{4}{c|}{\textbf{Antisemitic phrases}}                                                                                   & \multicolumn{4}{c}{\textbf{Islamophobic phrases}}                                                                                   \\ \midrule
          \multicolumn{1}{l}{\textbf{Phrase}}            & \multicolumn{1}{l}{\textbf{\# Textual}} & \multicolumn{1}{l}{\textbf{\#Visual}} & \multicolumn{1}{l|}{\textbf{Ratio}} &  \multicolumn{1}{l}{\textbf{Phrase}} & \multicolumn{1}{l}{\textbf{\#Textual}} & \multicolumn{1}{l}{\textbf{\#Visual }} & \multicolumn{1}{l}{\textbf{Ratio}} \\ 
          
          \midrule

kill yourself you filthy kike	&16&	0	&$\infty$& fuck off mudslime& 225 & 33 &6.8\\
fuck this kike shit	&12	&0	&$\infty$ & kill muslim & 4,716&1,303&3.6 \\
a kike&	51,216&	1& 51,216.0& fuck mudslimes	&985&491	&2.0\\
kill yourself kike	&1,059	&3	&353&fuck sandniggers	&895	&521	&1.7\\
i hate kike	&612	&2	&306&kill all muslim&1,733&1,528&1.1\\
\midrule
fuck off dirty jew	&9	&72,033	&0.00012& fuck off muslim shill	&12&	12,485&	0.00096 \\
fuck soros and fuck jewish people	&5	&45,805& 0.00011&	fuck muslim though	&11	&12,704&	0.00087\\
jewish child should be kill	&1	&9,699	&0.00010& all fuck muslim must fuck hang	&7	&10,170	&0.00069\\
fuck off jew scum&	7	&72,763 &	0.00010&fuck muhammad and fuck muslim	&7	&13,347	&0.00052\\
hitler didnt kill enough jew	&1	&36,621	&0.00003&fuck islam and fuck all muslim	&5&	12,732&	0.00039\\
\bottomrule
\end{tabular}
}
\caption{Antisemitic/Islamophobic phrases with the highest and lowest ratio (\#Posts in Textual/\#Posts in Visual).}
\label{tab:relevant_phrases_ratio_textual_visual_posts}
\end{table*}


In order to identify how many different images are posted, we calculate the CDF of the number of unique images per Antisemitic/Islamophobic phrase (see Fig.~\ref{fig:number_unique_images_per_phrase}). We find some differences between the CDF for Antisemitic and Islamophobic phrases. While 80,37\% of the Antisemitic phrases are associated with at least 100 different images, the percentage is higher (98.94\%) for the Islamophobic ones. In the same light, while 54.6\% of the Antisemitic phrases are connected with at least 1000 different images, 70.21\% of the Islamophobic phrases are. These results suggest that the diversity in imagery is higher for Islamophobic content on 4chan's /pol/. 




\begin{figure}[!t]
\centering
\includegraphics[width=0.8\columnwidth]{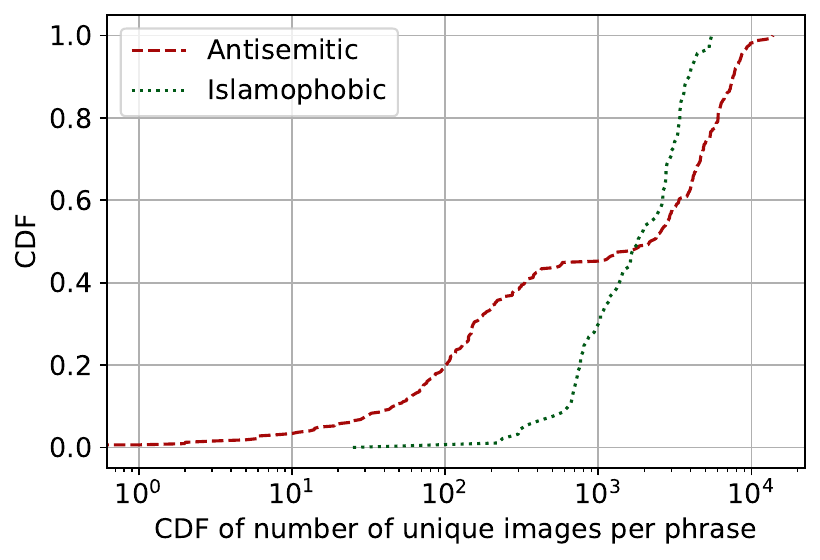}
\caption{CDF of the number of unique images per phrase.}
\label{fig:number_unique_images_per_phrase}
\end{figure}
}

\section{Related Work}

\descr{Hate Speech detection.} Hate speech has recently received much research attention, with several works focusing on detecting hate speech in online social media. Initial research on hate speech analysis is typically oriented towards monolingual and single classification tasks due to the complexity of the task. They used simple methods such as  dictionary lookup~\cite{4618794}, bag of words~\cite{4618794}, or SVM classifiers ~\cite{malmasi-zampieri-2017-detecting,9031482}. Recent efforts are proposing multilingual and multitask learning by using deep learning models~\cite{wang-etal-2020-detect, ousidhoum-etal-2019-multilingual,glavas-etal-2020-xhate,10.1145/3447535.3462495}. While previous approaches to characterize and identify hate speech focus purely on the \textit{content} posted in social media, some research efforts shift the focus towards detecting hateful users by exploiting other contextual data ~\cite{ribeiro2018characterizing,DBLP:journals/corr/abs-2103-11806,chaudhry2020you,waseem-hovy-2016-hateful}. Furthermore, other research efforts investigate to what extend the models trained to detect general abusive language generalize between different datasets labeled with different abusive language types~\cite{karan-snajder-2018-cross, meyer-gamback-2019-platform,rizoiu2019transfer,salminen2020developing,nejadgholi2020cross}. While less explored, some work focus on multimodal settings, formed by text and images~\cite{das2020detecting,kiela2020hateful}. ~\citet{gomez2020exploring} build  a large dataset for multimodal hate speech detection retrieved from Twitter using specific hateful seed keywords, finding that multimodal models do not  outperform the unimodal text ones.



\descr{Antisemitism.} Antisemitism have grown and proliferated rapidly online and have done so mostly unchecked; ~\citet{zannettou2020quantitative} call for new techniques to understand it better and combat it. \citet{doi:10.1177/2056305120916850} train a scalable supervised machine learning classifier to identify antisemitic content on Twitter. 
~\citet{chandra2021subverting} propose a multimodal system that uses text, images, and OCR to detect the presence of Antisemitic textual and visual content. They apply their model on Twitter and Gab, finding that multiple screenshots, multi-column text, and texts expressing irony, sarcasm posed problems for the classifiers. To characterize antisemitism, ~\citet{enstad2021contemporary} propose an analytical framework composed of three indicators: antisemitic attitudes, incidents targeting Jews, and Jew's exposure to antisemitism. Their results show that attitudes vary by geographic and cultural region and among population sub-groups. 

\descr{Islamophobia.} Surveys show that Islamophobia is rising on Web communities ~\cite{hafez2019european}. ~\citet{vidgen2020detecting} build an SVM classifier to distinguish between tweets non-Islamophobic, weak Islamophobic, and strong Islamophobic with a balanced accuracy of 83\%. ~\citet{cervi2020exclusionary}  use clause-based semantic text analysis to identify the presence of Islamophobia in electoral discourses of political parties from Spain and Italy. ~\citet{10.1145/3465336.3475111} apply topic modeling and temporal analysis over tweets from the \#coronajihad to identify the existence of Islamophobic rhetoric around COVID-19 in India. ~\citet{civila2020demonization} apply content analysis over 474  images and texts in  from Instagram  posts under the hashtag \#StopIslam. ~\citet{alietti2013religious} conduct telephone surveys on 1.5K Italians on Antisemitic and Islamophobic attitudes, finding an overlap of ideology for both types of hate speech. 

\section{Discussion \& Conclusion}

In this work, we explored the problem of Antisemitism/Islamophobia on 4chan's /pol/ using OpenAI's CLIP model.
We devised a methodology to identify Antisemitic/Islamophobic textual phrases using Google's Perspective API and manual annotations and then used the CLIP model to identify hateful imagery based on the phrases.
We found that the CLIP can play a role in detecting hateful content; using our methods, the CLIP can detect hateful content with an accuracy of 81\%.
Also, we found that Antisemitic/Islamophobic imagery exists in a similar number of posts when compared to Antisemitic/Islamophobic speech on 4chan's /pol/.
Additionally, our work contributes to research efforts focusing on understanding and detecting hateful content by making a dataset of 420 Antisemitic/Islamophobic phrases, 246K textual posts, and 21K images available (upon request).
Below, we discuss the implications of our findings for researchers focusing on detecting hate speech and for researchers working on large pre-trained models like OpenAI's CLIP.

\descr{Prevalence of Antisemitic/Islamophobic Imagery.} Our findings show that images play a significant role in the spread of hateful content. 
This is likely because 4chan is an imageboard and a fringe Web community; hence, a large volume of hateful content is disseminated via images.
Nevertheless, the problem of hateful imagery exists on other mainstream platforms (e.g., Twitter), hence it is of paramount importance to develop better and more accurate systems for the detection of hateful content across multiple modalities.
For instance, we argue that the spread of hateful content via videos is an unexplored problem, and there is a need to develop models across text, images, and videos.

\descr{Performance and Sensitivity of CLIP model.} Our experiments indicate that large-pre-trained models like CLIP are pretty powerful and have general knowledge that can be used for various tasks.
When considering the hateful content detection task, the CLIP model should be used with caution.
This is because the CLIP model highly depends on how the input text query is written, influencing the number of false positives returned.
When CLIP is used for moderation purposes, we argue that it is essential to have humans in the loop to ensure that the automated model works as expected.
\revision{Additionally, we observed that the CLIP model performs worse when considering input text queries that comprise many words. This poor performance also occurs when images contain text that are long (from our annotations we observed that many false positives are screenshots of images with a lot of text).} \revisioncomment{R3.1}.
This indicates that we need more powerful text encoders that can capture the primary meaning of textual phrases, irrespectively of how long they are.
\revision{Also, we emphasize that CLIP's  performance on detecting hateful content yields a substantial number of false positives, which is expected given the nature of the problem (i.e., hate speech is sometimes hard to identify/subjective). The same applies to all the baselines that we experimented with, in particular, to a larger extent since their precision score is poorer compared to CLIP. 
CLIP's poor precision score (0.54) is also reflected in the dataset that we are releasing. Researchers that aim to use the dataset for other downstream tasks (e.g., implementing classifiers for hateful content) should have this limitation in mind and potentially make additional manual annotations to decrease the number of false positives in the dataset.}\revisioncomment{R3.3}

\descr{Biases on CLIP model.} Large pre-trained models like OpenAI's CLIP are trained on large-scale datasets from the Web, and these datasets might include biases, hence some of the bias is transferred to the trained model.
From our experiments and manual annotations, we observed some instances of such biases; e.g., the CLIP model identifying an image showing a terrorist as similar to a text phrase talking about Muslims (i.e., the model is biased towards Muslims, thinking they are terrorists).
When considering that these models can be used for moderation purposes (e.g., detecting and removing hateful content), such biases can result in false positives biased towards specific demographics.
This can cause users to lose trust in the platform and its moderation systems and may cause them to stop using the platform.
Overall, given the increasing use of such models in real-world applications, there is a pressing need to develop techniques and tools to diminish such biases from large pre-trained models.

\descr{Limitations.} Our work has several limitations. First, we rely on Google's Perspective API to initially identify hateful text, which has its limitations (e.g., might not understand specific slurs posted on 4chan) and biases when detecting hateful text.
\revision{Second, our analysis focuses on a small number of short textual phrases (at most seven words), mainly because our preliminary results showed that CLIP does not perform well in detecting hateful imagery when considering long phrases. Therefore, our approach is likely to miss some Antisemitic/Islamophobic text and imagery because of the small number of phrases that we consider.}\revisioncomment{R2.1, R2.14, R2.5, R2.15}
Third, we rely entirely on a pre-trained CLIP model; this is not ideal since the CLIP model is trained on a public dataset obtained from multiple Web resources and is not specific to our platform of interest (i.e., 4chan). This might result in the model not recognizing some 4chan slurs or slang language.
\revision{Fourth, our work and analysis focuses on a single data source (4chan's /pol/), a limitation that does not allow us to investigate the performance of CLIP on other more mainstream communities like Twitter, Facebook, and Reddit (e.g., how CLIP performs on content shared on mainstream platforms).
As part of our future work, we plan to investigate CLIP's performance on other platform and we intend to fine-tune the CLIP model with datasets obtained from mainstream social networks such as Twitter and Reddit as well as fringe Web communities that are often associated with the alt-right (e.g., Gab). 
Finally, as discussed above, the released dataset includes a substantial number of false positives, which indicates that researchers should consider the existence of false positives in the dataset when using it.}
\revisioncomment{R2.2,MR.6}






\small

\bibliographystyle{plainnat}

\begin{thebibliography}{81}
\providecommand{\natexlab}[1]{#1}
\providecommand{\url}[1]{\texttt{#1}}
\providecommand{\urlprefix}{URL }
\expandafter\ifx\csname urlstyle\endcsname\relax
  \providecommand{\doi}[1]{doi:\discretionary{}{}{}#1}\else
  \providecommand{\doi}{doi:\discretionary{}{}{}\begingroup
  \urlstyle{rm}\Url}\fi

\bibitem[{Ahmed, Vidgen, and Hale(2021)}]{DBLP:journals/corr/abs-2103-11806}
Ahmed, Z.; Vidgen, B.; and Hale, S.~A. 2021.
\newblock Tackling Racial Bias in Automated Online Hate Detection: Towards Fair
  and Accurate Classification of Hateful Online Users Using Geometric Deep
  Learning.
\newblock \emph{CoRR} abs/2103.11806.

\bibitem[{Alietti and Padovan(2013)}]{alietti2013religious}
Alietti, A.; and Padovan, D. 2013.
\newblock Religious racism. Islamophobia and antisemitism in Italian society.
\newblock \emph{Religions} 4(4): 584--602.

\bibitem[{Anonymous(2022)}]{url_toxic_phrases}
Anonymous. 2022.
\newblock Common Antisemitic and Islamophobic phrases.
\newblock
  \url{https://docs.google.com/spreadsheets/d/1zgufPSiU8zHJd9E9bka8cagsBBCVNCM4ofLuGvgXG5w/edit?usp=sharing}.

\bibitem[{Arango, P{\'e}rez, and Poblete(2019)}]{arango2019hate}
Arango, A.; P{\'e}rez, J.; and Poblete, B. 2019.
\newblock Hate speech detection is not as easy as you may think: A closer look
  at model validation.
\newblock In \emph{SIGIR}.

\bibitem[{Baker, Haberman, and Thrush(2017)}]{baker2017trump}
Baker, P.; Haberman, M.; and Thrush, G. 2017.
\newblock Trump Removes Stephen Bannon From National Security Council Post.
\newblock
  \url{https://www.nytimes.com/2017/04/05/us/politics/national-security-council-stephen-bannon.html}.

\bibitem[{Bernstein et~al.(2011)Bernstein, Monroy-Hernandez, Harry, Andr,
  Panovich, and Vargas}]{bernstein2011analysis}
Bernstein, M.; Monroy-Hernandez, A.; Harry, D.; Andr, P.; Panovich, K.; and
  Vargas, G. 2011.
\newblock An Analysis of Anonymity and Ephemerality in a Large Online
  Community. ICWSM, 50--57.

\bibitem[{Brown et~al.(2020)Brown, Mann, Ryder, Subbiah, Kaplan, Dhariwal,
  Neelakantan, Shyam, Sastry, Askell et~al.}]{brown2020language}
Brown, T.~B.; Mann, B.; Ryder, N.; Subbiah, M.; Kaplan, J.; Dhariwal, P.;
  Neelakantan, A.; Shyam, P.; Sastry, G.; Askell, A.; et~al. 2020.
\newblock Language models are few-shot learners.
\newblock \emph{arXiv preprint arXiv:2005.14165} .

\bibitem[{Caruana and Niculescu-Mizil(2006)}]{caruana2006empirical}
Caruana, R.; and Niculescu-Mizil, A. 2006.
\newblock An empirical comparison of supervised learning algorithms.
\newblock In \emph{ICML}.

\bibitem[{Cervi(2020)}]{cervi2020exclusionary}
Cervi, L. 2020.
\newblock Exclusionary Populism and Islamophobia: A Comparative Analysis of
  Italy and Spain.
\newblock \emph{Religions} 11(10): 516.

\bibitem[{Chandra et~al.(2021{\natexlab{a}})Chandra, Pailla, Bhatia,
  Sanchawala, Gupta, Shrivastava, and Kumaraguru}]{chandra2021subverting}
Chandra, M.; Pailla, D.; Bhatia, H.; Sanchawala, A.; Gupta, M.; Shrivastava,
  M.; and Kumaraguru, P. 2021{\natexlab{a}}.
\newblock “Subverting the Jewtocracy”: Online Antisemitism Detection Using
  Multimodal Deep Learning.
\newblock In \emph{WebSci}.

\bibitem[{Chandra et~al.(2021{\natexlab{b}})Chandra, Reddy, Sehgal, Gupta,
  Buduru, and Kumaraguru}]{10.1145/3465336.3475111}
Chandra, M.; Reddy, M.; Sehgal, S.; Gupta, S.; Buduru, A.~B.; and Kumaraguru,
  P. 2021{\natexlab{b}}.
\newblock "A Virus Has No Religion": Analyzing Islamophobia on Twitter During
  the COVID-19 Outbreak.
\newblock In \emph{HT}.

\bibitem[{Chaudhry and Lease(2020)}]{chaudhry2020you}
Chaudhry, P.; and Lease, M. 2020.
\newblock You Are What You Tweet: Profiling Users by Past Tweets to Improve
  Hate Speech Detection.
\newblock \emph{arXiv preprint arXiv:2012.09090} .

\bibitem[{Chen et~al.(2020)Chen, Kornblith, Norouzi, and
  Hinton}]{chen2020simple}
Chen, T.; Kornblith, S.; Norouzi, M.; and Hinton, G. 2020.
\newblock A simple framework for contrastive learning of visual
  representations.
\newblock In \emph{ICML}.

\bibitem[{Civila, Romero-Rodr{\'\i}guez, and
  Civila(2020)}]{civila2020demonization}
Civila, S.; Romero-Rodr{\'\i}guez, L.~M.; and Civila, A. 2020.
\newblock The Demonization of Islam through Social Media: A Case Study of\#
  Stopislam in Instagram.
\newblock \emph{Publications} 8(4): 52.

\bibitem[{Cobain et~al.(2017)Cobain, Perraudin, Morris, and
  Parveen}]{cobain2017salman}
Cobain, I.; Perraudin, F.; Morris, S.; and Parveen, N. 2017.
\newblock Salman Ramadan Abedi Named by Police as Manchester Arena
  Attacker.”.
\newblock \emph{The Guardian} .

\bibitem[{Costa and Phillip(2017)}]{costa2017stephen}
Costa, R.; and Phillip, A. 2017.
\newblock Stephen Bannon removed from National Security Council.
\newblock
  \url{https://www.washingtonpost.com/news/post-politics/wp/2017/04/05/steven-bannon-no-longer-a-member-of-national-security-council/}.

\bibitem[{Das, Wahi, and Li(2020)}]{das2020detecting}
Das, A.; Wahi, J.~S.; and Li, S. 2020.
\newblock Detecting Hate Speech in Multi-modal Memes.
\newblock \emph{arXiv preprint arXiv:2012.14891} .

\bibitem[{Davidson et~al.(2017)Davidson, Warmsley, Macy, and
  Weber}]{davidson2017automated}
Davidson, T.; Warmsley, D.; Macy, M.; and Weber, I. 2017.
\newblock Automated hate speech detection and the problem of offensive
  language.
\newblock In \emph{ICWSM}.

\bibitem[{Devlin et~al.(2018)Devlin, Chang, Lee, and
  Toutanova}]{devlin2018bert}
Devlin, J.; Chang, M.-W.; Lee, K.; and Toutanova, K. 2018.
\newblock Bert: Pre-training of deep bidirectional transformers for language
  understanding.
\newblock \emph{arXiv preprint arXiv:1810.04805} .

\bibitem[{Diba et~al.(2021)Diba, Sharma, Safdari, Lotfi, Sarfraz, Stiefelhagen,
  and Van~Gool}]{diba2021vi2clr}
Diba, A.; Sharma, V.; Safdari, R.; Lotfi, D.; Sarfraz, S.; Stiefelhagen, R.;
  and Van~Gool, L. 2021.
\newblock Vi2CLR: Video and Image for Visual Contrastive Learning of
  Representation.
\newblock In \emph{ICCV}, 1502--1512.

\bibitem[{Dictionaries(2021)}]{antisemitism_definition}
Dictionaries, O.~L. 2021.
\newblock anti-Semitism.
\newblock
  \url{https://www.oxfordlearnersdictionaries.com/definition/american_english/anti-semitism}.

\bibitem[{Downing, Gerwens, and Dron(2022)}]{downing2022tweeting}
Downing, J.; Gerwens, S.; and Dron, R. 2022.
\newblock Tweeting terrorism: Vernacular conceptions of Muslims and terror in
  the wake of the Manchester Bombing on Twitter.
\newblock \emph{CTS} 1--28.

\bibitem[{Enstad(2021)}]{enstad2021contemporary}
Enstad, J.~D. 2021.
\newblock Contemporary Antisemitism in Three Dimensions: A New Framework for
  Analysis.
\newblock \emph{SocArXiv} 28.

\bibitem[{Gao, Yao, and Chen(2021)}]{gao2021simcse}
Gao, T.; Yao, X.; and Chen, D. 2021.
\newblock SimCSE: Simple Contrastive Learning of Sentence Embeddings.
\newblock \emph{arXiv preprint arXiv:2104.08821} .

\bibitem[{Giorgi et~al.(2020)Giorgi, Nitski, Bader, and
  Wang}]{giorgi2020declutr}
Giorgi, J.~M.; Nitski, O.; Bader, G.~D.; and Wang, B. 2020.
\newblock Declutr: Deep contrastive learning for unsupervised textual
  representations.
\newblock \emph{arXiv preprint arXiv:2006.03659} .

\bibitem[{Glava{\v{s}}, Karan, and Vuli{\'c}(2020)}]{glavas-etal-2020-xhate}
Glava{\v{s}}, G.; Karan, M.; and Vuli{\'c}, I. 2020.
\newblock {XH}ate-999: Analyzing and Detecting Abusive Language Across Domains
  and Languages.
\newblock In \emph{COLING}.

\bibitem[{Gomez et~al.(2020)Gomez, Gibert, Gomez, and
  Karatzas}]{gomez2020exploring}
Gomez, R.; Gibert, J.; Gomez, L.; and Karatzas, D. 2020.
\newblock Exploring hate speech detection in multimodal publications.
\newblock In \emph{WACV}, 1470--1478.

\bibitem[{Goodfellow et~al.(2014)Goodfellow, Pouget-Abadie, Mirza, Xu,
  Warde-Farley, Ozair, Courville, and Bengio}]{goodfellow2014generative}
Goodfellow, I.; Pouget-Abadie, J.; Mirza, M.; Xu, B.; Warde-Farley, D.; Ozair,
  S.; Courville, A.; and Bengio, Y. 2014.
\newblock Generative adversarial nets.
\newblock \emph{Advances in neural information processing systems} 27.

\bibitem[{Google(2021)}]{perspective_library}
Google. 2021.
\newblock Perspective.
\newblock \url{https://www.perspectiveapi.com/}.

\bibitem[{Guermazi, Hammami, and Hamadou(2007)}]{4618794}
Guermazi, R.; Hammami, M.; and Hamadou, A.~B. 2007.
\newblock Using a Semi-automatic Keyword Dictionary for Improving Violent Web
  Site Filtering.
\newblock In \emph{SITIS 2007}, 337--344.

\bibitem[{Haberman, Peters, and Baker(2017)}]{haberman2017battle}
Haberman, M.; Peters, J.~W.; and Baker, P. 2017.
\newblock In battle for Trump’s heart and mind, it’s Bannon vs. Kushner.
\newblock
  \url{https://www.nytimes.com/2017/04/06/us/politics/stephen-bannon-white-house.html}.

\bibitem[{Hadsell, Chopra, and LeCun(2006)}]{hadsell2006dimensionality}
Hadsell, R.; Chopra, S.; and LeCun, Y. 2006.
\newblock Dimensionality reduction by learning an invariant mapping.
\newblock In \emph{CVPR}.

\bibitem[{Hafez et~al.(2019)Hafez, Bayrakli, Faytre, Easat-Daas, Younes,
  Kutuzova, Kar{\v{c}}i{\'c}, Emin, Me{\v{s}}ki{\'c}, Dizdarevic
  et~al.}]{hafez2019european}
Hafez, F.; Bayrakli, E.; Faytre, L.; Easat-Daas, A.; Younes, A.-E.; Kutuzova,
  N.; Kar{\v{c}}i{\'c}, H.; Emin, H.~A.; Me{\v{s}}ki{\'c}, N.~K.; Dizdarevic,
  S.~M.; et~al. 2019.
\newblock European Islamophobia Report 2019.
\newblock \url{https://setav.org/en/assets/uploads/2020/06/EIR_2019.pdf}.

\bibitem[{Hassani and Khasahmadi(2020)}]{hassani2020contrastive}
Hassani, K.; and Khasahmadi, A.~H. 2020.
\newblock Contrastive multi-view representation learning on graphs.
\newblock In \emph{ICML}.

\bibitem[{Hine et~al.(2017)Hine, Onaolapo, De~Cristofaro, Kourtellis,
  Leontiadis, Samaras, Stringhini, and Blackburn}]{hine2017kek}
Hine, G.; Onaolapo, J.; De~Cristofaro, E.; Kourtellis, N.; Leontiadis, I.;
  Samaras, R.; Stringhini, G.; and Blackburn, J. 2017.
\newblock Kek, cucks, and god emperor trump: A measurement study of 4chan’s
  politically incorrect forum and its effects on the web.
\newblock In \emph{ICWSM}.

\bibitem[{Karan and {\v{S}}najder(2018)}]{karan-snajder-2018-cross}
Karan, M.; and {\v{S}}najder, J. 2018.
\newblock Cross-Domain Detection of Abusive Language Online.
\newblock In \emph{ALW2}, 132--137. Brussels, Belgium: Association for
  Computational Linguistics.

\bibitem[{Kiela et~al.(2019)Kiela, Bhooshan, Firooz, Perez, and
  Testuggine}]{kiela2019supervised}
Kiela, D.; Bhooshan, S.; Firooz, H.; Perez, E.; and Testuggine, D. 2019.
\newblock Supervised multimodal bitransformers for classifying images and text.
\newblock \emph{arXiv preprint arXiv:1909.02950} .

\bibitem[{Kiela et~al.(2020)Kiela, Firooz, Mohan, Goswami, Singh, Ringshia, and
  Testuggine}]{kiela2020hateful}
Kiela, D.; Firooz, H.; Mohan, A.; Goswami, V.; Singh, A.; Ringshia, P.; and
  Testuggine, D. 2020.
\newblock The hateful memes challenge: Detecting hate speech in multimodal
  memes.
\newblock \emph{arXiv preprint arXiv:2005.04790} .

\bibitem[{Kim et~al.(2020)Kim, Lee, Bae, and Yun}]{kim2020mixco}
Kim, S.; Lee, G.; Bae, S.; and Yun, S.-Y. 2020.
\newblock MixCo: Mix-up Contrastive Learning for Visual Representation.
\newblock \emph{arXiv preprint arXiv:2010.06300} .

\bibitem[{Konikoff(2021)}]{konikoff2021gatekeepers}
Konikoff, D. 2021.
\newblock Gatekeepers of toxicity: Reconceptualizing Twitter's abuse and hate
  speech policies.
\newblock \emph{P\&I} .

\bibitem[{Kvalseth(1989)}]{kappa}
Kvalseth, T.~O. 1989.
\newblock Note on Cohen's Kappa.
\newblock \emph{Psychol. Rep.} .

\bibitem[{Mahalanobis(1936)}]{mahalanobis1936generalized}
Mahalanobis, P.~C. 1936.
\newblock On the generalized distance in statistics.
\newblock National Institute of Science of India.

\bibitem[{Malmasi and Zampieri(2017)}]{malmasi-zampieri-2017-detecting}
Malmasi, S.; and Zampieri, M. 2017.
\newblock Detecting Hate Speech in Social Media.
\newblock In \emph{RANLP}.

\bibitem[{McInnes, Healy, and Melville(2018)}]{mcinnes2018umap}
McInnes, L.; Healy, J.; and Melville, J. 2018.
\newblock Umap: Uniform manifold approximation and projection for dimension
  reduction.
\newblock \emph{arXiv preprint arXiv:1802.03426} .

\bibitem[{Merriam-Webster(2021)}]{islamophobia_definition}
Merriam-Webster. 2021.
\newblock Islamophobia.
\newblock \url{https://www.merriam-webster.com/dictionary/Islamophobia}.

\bibitem[{Meyer and Gamb{\"a}ck(2019)}]{meyer-gamback-2019-platform}
Meyer, J.~S.; and Gamb{\"a}ck, B. 2019.
\newblock A Platform Agnostic Dual-Strand Hate Speech Detector.
\newblock In \emph{ALW}, 146--156. Florence, Italy: Association for
  Computational Linguistics.

\bibitem[{Monga and Evans(2006)}]{Monga2006PerceptualIH}
Monga, V.; and Evans, B.~L. 2006.
\newblock Perceptual Image Hashing Via Feature Points: Performance Evaluation
  and Tradeoffs.
\newblock \emph{TIP} 15: 3452--3465.

\bibitem[{Nejadgholi and Kiritchenko(2020)}]{nejadgholi2020cross}
Nejadgholi, I.; and Kiritchenko, S. 2020.
\newblock On cross-dataset generalization in automatic detection of online
  abuse.
\newblock \emph{arXiv preprint arXiv:2010.07414} .

\bibitem[{NLTK(2021{\natexlab{a}})}]{wordnet}
NLTK. 2021{\natexlab{a}}.
\newblock NLTK Lemmatization.
\newblock \url{https://www.nltk.org/_modules/nltk/stem/wordnet.html}.

\bibitem[{NLTK(2021{\natexlab{b}})}]{sentence_tokenizer}
NLTK. 2021{\natexlab{b}}.
\newblock NLTK Tokenization.
\newblock \url{https://www.nltk.org/api/nltk.tokenize.html}.

\bibitem[{Ousidhoum et~al.(2019)Ousidhoum, Lin, Zhang, Song, and
  Yeung}]{ousidhoum-etal-2019-multilingual}
Ousidhoum, N.; Lin, Z.; Zhang, H.; Song, Y.; and Yeung, D.-Y. 2019.
\newblock Multilingual and Multi-Aspect Hate Speech Analysis.
\newblock In \emph{EMNLP-IJCNLP}, 4675--4684. Hong Kong, China: Association for
  Computational Linguistics.

\bibitem[{Ozalp et~al.(2020)Ozalp, Williams, Burnap, Liu, and
  Mostafa}]{doi:10.1177/2056305120916850}
Ozalp, S.; Williams, M.~L.; Burnap, P.; Liu, H.; and Mostafa, M. 2020.
\newblock Antisemitism on Twitter: Collective Efficacy and the Role of
  Community Organisations in Challenging Online Hate Speech.
\newblock \emph{Soc. Media Soc.} .

\bibitem[{Papasavva et~al.(2020)Papasavva, Zannettou, De~Cristofaro,
  Stringhini, and Blackburn}]{papasavva2020raiders}
Papasavva, A.; Zannettou, S.; De~Cristofaro, E.; Stringhini, G.; and Blackburn,
  J. 2020.
\newblock Raiders of the lost kek: 3.5 years of augmented 4chan posts from the
  politically incorrect board.
\newblock In \emph{ICWSM}.

\bibitem[{{Perspective API}(2018)}]{jigsaw2018perspective}
{Perspective API}. 2018.
\newblock \url{https://www.perspectiveapi.com/}.

\bibitem[{Pramanick et~al.(2021)Pramanick, Sharma, Dimitrov, Akhtar, Nakov, and
  Chakraborty}]{pramanick-etal-2021-momenta-multimodal}
Pramanick, S.; Sharma, S.; Dimitrov, D.; Akhtar, M.~S.; Nakov, P.; and
  Chakraborty, T. 2021.
\newblock {MOMENTA}: A Multimodal Framework for Detecting Harmful Memes and
  Their Targets.
\newblock In \emph{Findings of the Association for Computational Linguistics:
  EMNLP 2021}, 4439--4455. Punta Cana, Dominican Republic: Association for
  Computational Linguistics.
\newblock \doi{10.18653/v1/2021.findings-emnlp.379}.
\newblock \urlprefix\url{https://aclanthology.org/2021.findings-emnlp.379}.

\bibitem[{Prisk(2017)}]{prisk2017hyperreality}
Prisk, D. 2017.
\newblock The hyperreality of the Alt Right: how meme magic works to create a
  space for far right politics. .

\bibitem[{Radford et~al.(2021)Radford, Kim, Hallacy, Ramesh, Goh, Agarwal,
  Sastry, Askell, Mishkin, Clark et~al.}]{radford2021learning}
Radford, A.; Kim, J.~W.; Hallacy, C.; Ramesh, A.; Goh, G.; Agarwal, S.; Sastry,
  G.; Askell, A.; Mishkin, P.; Clark, J.; et~al. 2021.
\newblock Learning transferable visual models from natural language
  supervision.
\newblock \emph{arXiv preprint arXiv:2103.00020} .

\bibitem[{Ribeiro et~al.(2018)Ribeiro, Calais, Santos, Almeida, and
  Meira~Jr}]{ribeiro2018characterizing}
Ribeiro, M.~H.; Calais, P.~H.; Santos, Y.~A.; Almeida, V.~A.; and Meira~Jr, W.
  2018.
\newblock Characterizing and detecting hateful users on twitter.
\newblock In \emph{ICWSM}.

\bibitem[{Ribeiro et~al.(2021)Ribeiro, Jhaver, Zannettou, Blackburn,
  De~Cristofaro, Stringhini, and West}]{ribeiro2020does}
Ribeiro, M.~H.; Jhaver, S.; Zannettou, S.; Blackburn, J.; De~Cristofaro, E.;
  Stringhini, G.; and West, R. 2021.
\newblock Does Platform Migration Compromise Content Moderation? Evidence from
  r/The\_Donald and r/Incels.
\newblock In \emph{CSCW}.

\bibitem[{Rivers and Lewis(2014)}]{rivers2014ethical}
Rivers, C.~M.; and Lewis, B.~L. 2014.
\newblock Ethical research standards in a world of big data.
\newblock \emph{F1000Research} 3.

\bibitem[{Rizoiu et~al.(2019)Rizoiu, Wang, Ferraro, and
  Suominen}]{rizoiu2019transfer}
Rizoiu, M.-A.; Wang, T.; Ferraro, G.; and Suominen, H. 2019.
\newblock Transfer learning for hate speech detection in social media.
\newblock \emph{arXiv preprint arXiv:1906.03829} .

\bibitem[{Rosenfeld(2017)}]{rosenfeld2017trump}
Rosenfeld, E. 2017.
\newblock Trump launches attack on Syria with 59 Tomahawk missiles.
\newblock
  \url{https://www.cnbc.com/2017/04/06/us-military-has-launched-more-50-than-missiles-aimed-at-syria-nbc-news.html}.

\bibitem[{Salminen et~al.(2020)Salminen, Hopf, Chowdhury, Jung, Almerekhi, and
  Jansen}]{salminen2020developing}
Salminen, J.; Hopf, M.; Chowdhury, S.~A.; Jung, S.-g.; Almerekhi, H.; and
  Jansen, B.~J. 2020.
\newblock Developing an online hate classifier for multiple social media
  platforms.
\newblock \emph{HCIS} 10(1): 1--34.

\bibitem[{Schuhmann et~al.(2021)Schuhmann, Vencu, Beaumont, Kaczmarczyk,
  Mullis, Katta, Coombes, Jitsev, and Komatsuzaki}]{schuhmann2021laion}
Schuhmann, C.; Vencu, R.; Beaumont, R.; Kaczmarczyk, R.; Mullis, C.; Katta, A.;
  Coombes, T.; Jitsev, J.; and Komatsuzaki, A. 2021.
\newblock Laion-400m: Open dataset of clip-filtered 400 million image-text
  pairs.
\newblock \emph{arXiv preprint arXiv:2111.02114} .

\bibitem[{Sellars(2016)}]{sellars2016defining}
Sellars, A. 2016.
\newblock Defining hate speech.
\newblock \emph{Berkman Klein Center Research Publication} (2016-20): 16--48.

\bibitem[{Senarath and Purohit(2020)}]{9031482}
Senarath, Y.; and Purohit, H. 2020.
\newblock Evaluating Semantic Feature Representations to Efficiently Detect
  Hate Intent on Social Media.
\newblock In \emph{2020 IEEE 14th International Conference on Semantic
  Computing (ICSC)}, 199--202.

\bibitem[{Tuters, Jokubauskait{\.e}, and Bach(2018)}]{tuters2018post}
Tuters, M.; Jokubauskait{\.e}, E.; and Bach, D. 2018.
\newblock Post-truth protest: how 4chan cooked up the Pizzagate Bullshit.
\newblock \emph{M/c Journal} 21(3).

\bibitem[{{Urban Dictionary}(2006{\natexlab{a}})}]{mudslime}
{Urban Dictionary}. 2006{\natexlab{a}}.
\newblock Mudslime definition.
\newblock \url{https://www.urbandictionary.com/define.php?term=mudslime}.

\bibitem[{{Urban Dictionary}(2006{\natexlab{b}})}]{sandnigger}
{Urban Dictionary}. 2006{\natexlab{b}}.
\newblock Sandniggers definition.
\newblock
  \url{https://www.urbandictionary.com/define.php?term=sand\%20niggers}.

\bibitem[{Vidgen and Yasseri(2020)}]{vidgen2020detecting}
Vidgen, B.; and Yasseri, T. 2020.
\newblock Detecting weak and strong Islamophobic hate speech on social media.
\newblock \emph{J. Inf. Technol. Politics} 17(1): 66--78.

\bibitem[{Vitiugin, Senarath, and Purohit(2021)}]{10.1145/3447535.3462495}
Vitiugin, F.; Senarath, Y.; and Purohit, H. 2021.
\newblock Efficient Detection of Multilingual Hate Speech by Using Interactive
  Attention Network with Minimal Human Feedback.
\newblock In \emph{WebSci}.

\bibitem[{Wang et~al.(2020)Wang, Lu, Han, Long, and
  Poon}]{wang-etal-2020-detect}
Wang, K.; Lu, D.; Han, C.; Long, S.; and Poon, J. 2020.
\newblock Detect All Abuse! Toward Universal Abusive Language Detection Models.
\newblock In \emph{COLING}, 6366--6376. Barcelona, Spain (Online):
  International Committee on Computational Linguistics.

\bibitem[{Waseem and Hovy(2016)}]{waseem-hovy-2016-hateful}
Waseem, Z.; and Hovy, D. 2016.
\newblock Hateful Symbols or Hateful People? Predictive Features for Hate
  Speech Detection on {T}witter.
\newblock In \emph{NAACL}, 88--93. San Diego, California: Association for
  Computational Linguistics.

\bibitem[{Wu et~al.(2020)Wu, Wang, Gu, Khabsa, Sun, and Ma}]{wu2020clear}
Wu, Z.; Wang, S.; Gu, J.; Khabsa, M.; Sun, F.; and Ma, H. 2020.
\newblock Clear: Contrastive learning for sentence representation.
\newblock \emph{arXiv preprint arXiv:2012.15466} .

\bibitem[{You et~al.(2020)You, Chen, Sui, Chen, Wang, and Shen}]{you2020graph}
You, Y.; Chen, T.; Sui, Y.; Chen, T.; Wang, Z.; and Shen, Y. 2020.
\newblock Graph contrastive learning with augmentations.
\newblock \emph{Adv. Neural Inf. Process. Syst.} 33: 5812--5823.

\bibitem[{Yuan et~al.(2021)Yuan, Lin, Kuen, Zhang, Wang, Maire, Kale, and
  Faieta}]{yuan2021multimodal}
Yuan, X.; Lin, Z.; Kuen, J.; Zhang, J.; Wang, Y.; Maire, M.; Kale, A.; and
  Faieta, B. 2021.
\newblock Multimodal Contrastive Training for Visual Representation Learning.
\newblock In \emph{CVPR}.

\bibitem[{Zannettou et~al.(2018)Zannettou, Caulfield, Blackburn, De~Cristofaro,
  Sirivianos, Stringhini, and Suarez-Tangil}]{zannettou2018origins}
Zannettou, S.; Caulfield, T.; Blackburn, J.; De~Cristofaro, E.; Sirivianos, M.;
  Stringhini, G.; and Suarez-Tangil, G. 2018.
\newblock On the origins of memes by means of fringe web communities.
\newblock In \emph{IMC}.

\bibitem[{Zannettou et~al.(2017)Zannettou, Caulfield, De~Cristofaro,
  Kourtellis, Leontiadis, Sirivianos, Stringhini, and
  Blackburn}]{zannettou2017web}
Zannettou, S.; Caulfield, T.; De~Cristofaro, E.; Kourtellis, N.; Leontiadis,
  I.; Sirivianos, M.; Stringhini, G.; and Blackburn, J. 2017.
\newblock The web centipede: understanding how web communities influence each
  other through the lens of mainstream and alternative news sources.
\newblock In \emph{IMC}.

\bibitem[{Zannettou et~al.(2020{\natexlab{a}})Zannettou, ElSherief, Belding,
  Nilizadeh, and Stringhini}]{zannettou2020measuring}
Zannettou, S.; ElSherief, M.; Belding, E.; Nilizadeh, S.; and Stringhini, G.
  2020{\natexlab{a}}.
\newblock Measuring and characterizing hate speech on news websites.
\newblock In \emph{WebSci}.

\bibitem[{Zannettou et~al.(2020{\natexlab{b}})Zannettou, Finkelstein, Bradlyn,
  and Blackburn}]{zannettou2020quantitative}
Zannettou, S.; Finkelstein, J.; Bradlyn, B.; and Blackburn, J.
  2020{\natexlab{b}}.
\newblock A quantitative approach to understanding online antisemitism.
\newblock In \emph{ICWSM}.

\bibitem[{Zhang et~al.(2020)Zhang, Jiang, Miura, Manning, and
  Langlotz}]{zhang2020contrastive}
Zhang, Y.; Jiang, H.; Miura, Y.; Manning, C.~D.; and Langlotz, C.~P. 2020.
\newblock Contrastive learning of medical visual representations from paired
  images and text.
\newblock \emph{arXiv preprint arXiv:2010.00747} .

\end{thebibliography}

%
\end{document}